\documentclass[11pt,a4paper]{article}
\pdfoutput=1

\usepackage{jheppub}

\usepackage{amsmath,amsthm,amssymb,multirow,psfrag}

\usepackage{feynmp}                                     
\usepackage{hyperref,graphicx}           
\usepackage{epsfig}
\usepackage{color}
\usepackage{lipsum}

\def\lsim{\mathrel{\rlap{\lower4pt\hbox{\hskip1pt$\sim$}}
  \raise1pt\hbox{$<$}}}
\def\gsim{\mathrel{\rlap{\lower4pt\hbox{\hskip1pt$\sim$}}
  \raise1pt\hbox{$>$}}}

\newcommand{\beq}{\begin{equation}}
\newcommand{\eeq}{\end{equation}}
\newcommand{\bea}{\begin{eqnarray}}
\newcommand{\eea}{\end{eqnarray}}

\newcommand{\bwt}{\begin{widetext}}
\newcommand{\ewt}{\end{widetext}}

\newcommand{\cO}{{\cal O}}

\newcommand{\eref}[1]{Eq.~\eqref{eq:#1}}

\DeclareRobustCommand{\Sec}[1]{Sec.~\ref{#1}}

\DeclareRobustCommand{\App}[1]{App.~\ref{#1}}

\DeclareRobustCommand{\Fig}[1]{Fig.~\ref{#1}}

\DeclareRobustCommand{\Eq}[1]{Eq.~(\ref{#1})}

\DeclareRobustCommand{\Fig}[1]{Fig.~\ref{#1}}

\begin{document}

\title{Putting a Stop to di-Higgs Modifications}

\author[a]{Brian Batell,}
\author[a]{Matthew McCullough,}
\author[a]{Daniel Stolarski,}
\author[b]{ and Christopher B. Verhaaren,}

\affiliation[a]{Theory Division, Physics Department, CERN, CH-1211 Geneva 23, Switzerland}
\affiliation[b]{Maryland Center for Fundamental Physics, Department of Physics, University of Maryland, College Park, MD 20742-4111}

\emailAdd{brian.batell@cern.ch}
\emailAdd{matthew.mccullough@cern.ch}
\emailAdd{daniel.stolarski@cern.ch}
\emailAdd{cver@umd.edu}

\abstract{
Pair production of Higgs bosons at hadron colliders is an enticing channel to search for new physics. New colored particles that couple strongly to the Higgs, such as those most often called upon to address the hierarchy problem, provide well motivated examples in which large enhancements of the di-Higgs rate are possible, at least in principle. However, in such scenarios the di-Higgs production rate is tightly correlated with the single Higgs production rate and, since the latter is observed to be SM-like, one generally expects that only modest enhancements in di-Higgs production are allowed by the LHC Run 1 data. We examine the contribution of top squarks (stops) in a simplified supersymmetry model to di-Higgs production and find that this general expectation is indeed borne out. In particular, the allowed deviations are typically small, but there are tuned regions of parameter space where expectations based on EFT arguments break down in which ${\cal O}(100\%)$ enhancements to the di-Higgs production rate are possible and are simultaneously consistent with the observed single Higgs production rates. These effects are potentially observable with the high luminosity run of the LHC or at a future hadron collider. 
}

\preprint{CERN-PH-TH-2015-186}
\maketitle

\section{Introduction} 
\label{sec:intro} 

A comprehensive experimental program to characterize the 125 GeV Higgs boson~\cite{Aad:2012tfa,Chatrchyan:2012ufa} and determine the underlying nature of electroweak symmetry breaking is underway at the LHC. 
Based on the complete Run 1 data set, significant progress has been made through the study of final states with a single Higgs particle. The largest and best measured single Higgs production channel is the one loop gluon fusion process, which is in good agreement with the predictions of the Standard Model (SM)~\cite{Khachatryan:2014jba,ATLAS-CONF-2015-007}. In addition, an important long term goal of this program is to observe and study final states with two Higgs bosons. The di-Higgs channel 
is sensitive to the trilinear self coupling of the Higgs particle, which in turn gives information about the shape of the scalar potential, and can furthermore provide a sensitive probe of physics beyond the SM (BSM). 

 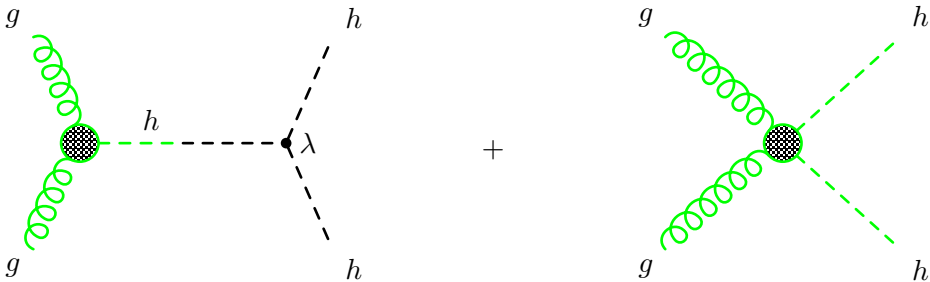
\begin{figure}[th]
  \vspace{5mm}
\begin{tabular}{ccc}
    \begin{tabular}{c}
\begin{fmffile}{dihiggsEFTtri}
\begin{fmfgraph*}(140,80)
\fmfpen{1.0}
\fmfleft{i1,p1,i2} \fmfright{o1,p2,o2}
\fmfv{l=$g$}{i1}\fmfv{l=$g$}{i2}
\fmfv{l=$h$}{o1}\fmfv{l=$h$}{o2}
\fmf{gluon,foreground=green}{v1,i1}\fmf{gluon,foreground=green}{v1,i2} 
\fmf{dashes,tension=0.9,foreground=green}{v1,v2}\fmf{dashes,tension=0.9}{v2,v3}
\fmf{dashes}{v3,o1}\fmf{dashes}{v3,o2}
\fmfv{decor.shape=circle,decor.filled=gray50,decor.size=7thick,l=$h$,l.a=20,l.d=25,foreground=green}{v1} \fmfv{decor.shape=circle,decor.filled=full,decor.size=1.5thick,l=$\lambda$,l.a=0,l.d=5}{v3}
\end{fmfgraph*}
\end{fmffile}
\end{tabular} &
\begin{tabular}{c}
\hspace{5mm}$+$ \hspace{5mm}
\end{tabular} &
\begin{tabular}{c}
\begin{fmffile}{dihiggsEFTquad}
\begin{fmfgraph*}(110,80)
\fmfpen{1.0}
\fmfleft{i1,p1,i2} \fmfright{o1,p2,o2}
\fmfv{l=$g$}{i1}\fmfv{l=$g$}{i2}
\fmfv{l=$h$}{o1}\fmfv{l=$h$}{o2}
\fmf{gluon,foreground=green}{v1,i1}\fmf{gluon,foreground=green}{v1,i2} 
\fmf{dashes,foreground=green}{v1,o1}\fmf{dashes,foreground=green}{v1,o2}
\fmfv{decor.shape=circle,decor.filled=gray50,decor.size=7thick,foreground=green}{v1}
\end{fmfgraph*}
\end{fmffile}
\end{tabular}
\end{tabular}
\caption{\label{fig:dihiggsEFTdiagrams} The relevant diagrams contributing to gluon fusion to di-Higgs with EFT vertices. The green lines indicate the core amplitudes focused on in this work. We refer to the diagrams on the left (right) as triangle (box) diagrams because of their topology in the SM.}
 \end{figure}

Like single Higgs production, the dominant di-Higgs production channel at the LHC is gluon fusion, which is depicted in \Fig{fig:dihiggsEFTdiagrams}.
In the SM and its extensions, di-Higgs production probes a different combination of couplings and masses than other loop processes such as single Higgs production via gluon fusion. One could then imagine that, even if for some reason the $hGG$ coupling were SM-like, there could be large deviations in di-Higgs production. This expectation is further motivated by the fact that in the SM the two diagrams\footnote{Throughout this work we refer to the diagrams on the left (right) of \Fig{fig:dihiggsEFTdiagrams} as triangle (box) diagrams because of their topology in the SM.} of \Fig{fig:dihiggsEFTdiagrams} interfere destructively making the SM di-Higgs production cross section smaller than the naive expectation~\cite{Li:2013rra,Dicus:2015yva,Dawson:2015oha}. Thus, typical BSM scenarios provide ample opportunity for significant modifications of di-Higgs production at hadron colliders when this cancellation is spoiled. Indeed this is the case for models with modified electroweak sectors and/or models where di-Higgs production may be resonantly enhanced through the production of new heavy fields which decay to Higgs pairs. 

In this work we instead focus on another potential source of modifications: new colored fields that couple to the Higgs. We investigate how much these scenarios may modify di-Higgs production at the LHC and future hadron colliders through their impact on the momentum-dependent $hGG$ and $h^2GG$ vertices (shaded green in \Fig{fig:dihiggsEFTdiagrams}) while keeping the Higgs quartic coupling $\lambda$ at its SM value. Throughout we refer to these as `non-resonant' corrections.

As a first step, consider the effective field theory (EFT) below some cutoff $\Lambda$ for the Higgs-gluon couplings $hGG$ and $h^2GG$. In general, if new heavy colored fields that couple to the Higgs and have mass $m\sim \Lambda$ are integrated out, they generate operators of the form
\begin{equation}
\left(\frac{c_1}{\Lambda^2}|H|^2+\frac{c_2}{\Lambda^4}|H|^4+\ldots \right)G_{\mu\nu}G^{\mu\nu}\, , \label{eq:decouplingOpshhgg}
\end{equation}
where $H$ is the Higgs doublet in the unbroken theory. In the broken theory, we can write the operators in terms of the physical Higgs field, $h$, and work to quadratic order in $h$. If we also include the SM contribution, which we will denote with the coefficient $c_{\text{SM}}\simeq\alpha_s/12\pi$, we obtain the effective operators
\begin{equation}
\frac{h}{\sqrt{2} v} \left(c_{\text{SM}}+\frac{2 c_1 v^2}{\Lambda^2}+\frac{4 c_2 v^4}{\Lambda^4}+\ldots \right) G_{\mu\nu}G^{\mu\nu}
+\frac{h^2}{4v^2}\left(-c_{\text{SM}}+\frac{2 c_1 v^2}{\Lambda^2} +\frac{12 c_2 v^4}{\Lambda^4}+\ldots \right)G_{\mu\nu}G^{\mu\nu} ,
\label{eq:EFT}
\end{equation}
where $v = 174$ GeV and the sign flip between single and double Higgs couplings in the SM has been included.

We now introduce a core observation from the first run of the LHC: modifications to the total single Higgs production rate are small. For a model in which the only BSM physics is new colored fields coupled to the Higgs, the cross section modifications must be $\lesssim \mathcal{O}(20 \%)$ \cite{Khachatryan:2014jba,ATLAS-CONF-2015-007}, implying modifications to the $hGG$ coupling of $\lesssim \mathcal{O}(10 \%)$. We may understand the implications of this observation for non-resonant contributions to di-Higgs production by studying \Eq{eq:EFT} more closely. 

If the new physics is heavy and respects decoupling, the usual rules of EFT apply. In particular, small corrections to single Higgs production imply $c_1 v^2/\Lambda^2 \ll c_{\text{SM}}$ and we can safely ignore the higher order terms. Then, \Eq{eq:EFT} implies that the magnitude of corrections to the $h^2 GG$ coupling must also be small if corrections to $hGG$ are small, as the magnitude of both are controlled by the same parameter combination $c_1 v^2/\Lambda^2 \ll c_{\text{SM}}$. Thus, we should expect non-resonant contributions in both diagrams for the di-Higgs production amplitude of \Fig{fig:dihiggsEFTdiagrams} to be similarly suppressed. 

It is worth noting that the impact of non-resonant new physics generically exhibits constructive interference between the triangle and box diagrams, unlike the top contribution in the SM. This implies that non-resonant corrections to di-Higgs production may spoil the cancellation in the SM and be larger than corrections to single Higgs production, but this will not be a very large effect. 

Quite generally then, the constraint that the $hGG$ coupling be SM-like implies that models with only colored, non-resonant, BSM states will have fairly SM-like di-Higgs rates for regions where the EFT is valid and current single Higgs constraints are taken into account. Clearly, the best chance for large deviations in the SM di-Higgs rate in this scenario is that the new particles are somewhat light so that an EFT analysis is inapplicable. In this case models must be checked on a case by case basis. In this work we explore this possibility in the context of scalar top partners (stops) in a simplified model as a supersymmetric extension of the SM. 

Supersymmetry is attractive because it provides a solution to the hierarchy problem. In a natural SUSY model one expects stops with masses below the TeV scale. Such stops have been searched for directly at colliders, but these searches depend strongly on the superpartner spectrum and specific decay modes of the stop. The bounds on stops decaying to a top and neutral LSP are approaching the TeV scale when the LSP is light~\cite{Chatrchyan:2013xna,Aad:2014kra,Aad:2014bva,CMS-PAS-SUS-14-011}, and are expected to get stronger with future LHC data~\cite{Stolarski:2013msa,Gershtein:2013iqa}. The bounds on very light stops, with masses in the 100 - 200 GeV range are much more difficult to evade. One possibility is that the stop could decay in a way that makes it much harder to discover at a collider. For example, it could be stealthy and nearly degenerate with the top~\cite{Fan:2011yu,Fan:2012jf,Csaki:2012fh,Han:2012fw,Kilic:2012kw,Czakon:2014fka}, or part of a compressed spectrum such that it is heavy but approximately degenerate with the particle it decays to~\cite{LeCompte:2011fh,LeCompte:2011cn,Dreiner:2012gx,Bhattacherjee:2012mz,Drees:2012dd,Belanger:2012mk,Alves:2012ft,Krizka:2012ah},\footnote{For recent models which predict such a compressed spectrum, see~\cite{Alves:2013wra,Dimopoulos:2014psa}.} or decay into other light MSSM particles (e.g. staus \cite{Carena:2012gp,Carena:2013iba}), or decay via baryon number $R$-parity violation~\cite{Brust:2012uf,Evans:2012bf,Bai:2013xla} where LHC searches are just starting to become sensitive~\cite{Khachatryan:2014lpa}. Because stops can be hidden in various exotic decay modes, complementary indirect bounds on top squarks are a crucial tool in the exploration of weak scale SUSY.

Indirect probes of stops include modifications to the $W$ mass~\cite{Barger:2012hr,Heinemeyer:2013dia}, corrections to Higgs production rates and branching ratios~\cite{Espinosa:2012in,Fan:2014txa} in loop processes, Higgs kinematic distributions~\cite{Grojean:2013nya,Schlaffer:2014osa} especially at high $p_T$, effects on Higgs wavefunction renormalization~\cite{Craig:2013xia,Gori:2013mia}, and stop-onium resonances~\cite{Martin:2008sv,Martin:2009dj,Younkin:2009zn,Batell:2015zla}. Stronger constraints could be obtained with future colliders~\cite{Craig:2014una,Fan:2014axa}. Because these probes of new physics are indirect, if a deviation is found it will be difficult to solve the inverse problem: what is the nature of the new physics that modifies a particular observable? Therefore, it is very important to explore as many different complementary probes as possible.

Higgs pair production has been studied in the Minimal Supersymmetric Standard Model (MSSM)~\cite{Plehn:1996wb,Djouadi:1999rca}, with~\cite{Belyaev:1999mx,BarrientosBendezu:2001di} exploring the effects of scalars in loops. In this paper we show, using stops as a concrete and well motivated example, that the absence of large deviations in single Higgs gluon fusion makes it very difficult to generate large enhancements in double Higgs production from non-resonant contributions alone. We show this in the context of an effective field theory and also with stops using low energy theorems~\cite{Shifman:1979eb,Kniehl:1995tn,Gillioz:2012se,Kribs:2012kz} as well as with a full loop calculation~\cite{Belyaev:1999mx,BarrientosBendezu:2001di}.
Despite these considerations, we do find that current Higgs data allow small, tuned, regions of parameter space with $\cO(1)$ deviations in the di-Higgs total cross section. 

In the following section, we survey the experimental and phenomenological literature on di-Higgs production at hadron colliders. While there is still significant uncertainty, we use it to select sensitivity benchmarks that we will use in this study. In Sec.~\ref{sec:EFT}, we analyze generic (and decoupling) heavy physics contributions to di-Higgs production using effective field theory, while in Sec.~\ref{sec:GenStopMod} we analyze heavy stops in the non-decoupling regime using low energy theorems. Finally in Sec.~\ref{sec:loop} we do a full loop calculation which is necessary for the case of light stops, and we find regions of parameter space where di-Higgs production has potentially observable modifications which are nonetheless consistent with single Higgs production constraints from Run 1. We conclude in Sec.~\ref{sec:conclusion}, and we give results for a 100 TeV collider in the appendix. 

\section{Collider Phenomenology}
\label{sec:ColPheno}

We begin by reviewing the prospects to measure the di-Higgs channel at the LHC and future hadron colliders. Due to its importance in understanding electroweak symmetry breaking, di-Higgs production is a well studied channel. 
In the SM the di-Higgs production rate was calculated long ago~\cite{Dicus:1987ez,Eboli:1987dy}, and at LHC energies the gluon fusion channel (see \Fig{fig:dihiggsEFTdiagrams}) dominates~\cite{Glover:1987nx}. This process was computed at leading order (LO)~\cite{Glover:1987nx,Plehn:1996wb} and next-to-leading order (NLO) in the heavy top limit~\cite{Dawson:1998py}, with more recent computations including higher orders in $1/m_t$~\cite{Grigo:2013rya,deFlorian:2013jea,Maltoni:2014eza}, parton shower effects~\cite{Frederix:2014hta}, and virtual corrections~\cite{Grigo:2014jma}. There are also computations of di-Higgs plus one jet~\cite{Dolan:2012rv,Li:2013flc,Maierhofer:2013sha} and vector boson fusion (di-Higgs plus two jets)~\cite{Dolan:2013rja}. The computations continue to improve, but due to the difficulty of the final state, the uncertainty in projecting the collider reach in this channel is dominated by experimental challenges.

With Run 1 data, ATLAS has released a search for non-resonant di-Higgs in the $bb\gamma\gamma$ channel~\cite{Aad:2014yja} setting a limit three orders of magnitude above the SM prediction.\footnote{This search sees a $2.4\sigma$ excess, but as we will see below, this excess cannot be explained by new particles running in loops.} There are also resonant searches in the $4b$ channel from CMS~\cite{CMS-PAS-HIG-14-013} and ATLAS~\cite{ATLAS-CONF-2014-005}, and in the $bb\gamma\gamma$~\cite{CMS-PAS-HIG-13-032} and the multi-lepton/photon channel~\cite{CMS-PAS-HIG-13-025} from CMS, all of which have cross section limits that are $\mathcal{O}({\rm pb})$, while the pair production cross section in the SM at 8 TeV is $\mathcal{O}({\rm fb})$. Future projections depend very strongly on the projections for experimental efficiencies and systematics. Preliminary studies for high luminosity LHC at ATLAS~\cite{ATL-PHYS-PUB-2014-019} and CMS~\cite{cms-talk} in the $bb\gamma\gamma$ channel and CMS in the $bbWW$~\cite{cms-talk} show a marginal sensitivity to observing pair production with 3,000 fb$^{-1}$ at 14 TeV, but further studies are ongoing.

There are also phenomenological studies that are more optimistic about the reach, but their sensitivity estimates vary greatly, even among those considering the same channels. For the most studied channel, $bb\gamma\gamma$~\cite{Baur:2003gp,Baglio:2012np,Yao:2013ika,Barger:2013jfa,Azatov:2015oxa,He:2015spf} significance estimates span from about $2\sigma$ to $6\sigma$. Other channels, including $bb\tau\tau$~\cite{Dolan:2012rv,Baglio:2012np,Barr:2013tda}, $bbWW$~\cite{Dolan:2012rv,Baglio:2012np,Papaefstathiou:2012qe}, and $4b$~\cite{Dolan:2012rv,deLima:2014dta,Wardrope:2014kya} have similar qualitative variance in the observability of these channels. Therefore, we take uncertainty benchmarks of 30\% and 60\% for observing deviations from the total SM rate, but ultimately more study will be needed to determine the true sensitivity of future searches.  

It is important to note, however, that di-Higgs modifications from stops will also lead to a modified spectrum in the di-Higgs invariant mass $m_{hh}$ or $p_T$. Thus, to obtain the strongest possible limit one would ideally perform an analysis which is sensitive to not only the total cross section but also the spectrum, especially features at higher center of mass energies. Such an analysis would depend heavily on the final state which is being observed. Therefore, instead of a full shape analysis for a specific final state we consider two invariant mass bins to demonstrate the importance of considering the spectrum.

If loops of new particles such as stops are responsible for a modification to the di-Higgs total rate, then other di-Higgs production channels will have SM-like rates and can be used to disentangle new physics scenarios. Vector boson fusion is a large component of di-Higgs plus two jets. This channel has been studied~\cite{Contino:2010mh,Baglio:2012np,Dolan:2013rja} but because of the small cross section, it is quite challenging at the LHC. Higgs pair production in association with $\bar{t}t$ is another challenging channel~\cite{Englert:2014uqa,Liu:2015aka}, but perhaps a combination of these channels in conjunction with improvements in collider analysis could yield sensitivity in the future. 
Di-Higgs production has also been explored for physics beyond the SM, both in the context of effective field theory~\cite{Pierce:2006dh,Goertz:2014qta,Azatov:2015oxa,Grober:2015cwa,Lu:2015jza,He:2015spf}, as well as for various specific new physics models~\cite{Plehn:1996wb,Djouadi:1999rca,Belyaev:1999mx,BarrientosBendezu:2001di,Arhrib:2009hc,Asakawa:2010xj,Kribs:2012kz,Dawson:2012mk,Dolan:2012ac,Cao:2013si,Han:2013sga,Nishiwaki:2013cma,Haba:2013xla,Enkhbat:2013oba,Chen:2014xra,Chen:2014xwa,Cao:2014kya,Chen:2014ask,vanBeekveld:2015tka,Dawson:2015oha,Wu:2015nba,Enkhbat:2015bca,Etesami:2015caa,Dall'Osso:2015aia,Lu:2015qqa}.

Planning is underway for higher energy hadron colliders where the cross section for Higgs pair production increases and prospects for measurements are potentially dramatically improved. The details of any putative collider and detector are still largely uncertain, but there have been several phenomenological studies of this process. The $bb\gamma\gamma$~\cite{Barr:2014sga,Azatov:2015oxa}, $4W$~\cite{Li:2015yia}, and $bb\,+$ leptons (and possibly also photon or missing energy)~\cite{Papaefstathiou:2015iba} all appear to be promising ways to measure di-Higgs production at a 100 TeV collider. In \App{sec:100TeV} we consider modifications to di-Higgs production due to stops for a 100 TeV proton-proton collider, taking precision benchmarks of 10\% and 20\% on the rate.

Finally, as we have emphasized, it is useful to compare the process $gg\rightarrow hh$ to $gg\rightarrow h$. The fitted rates for single Higgs production in gluon fusion, normalized to the SM value, are $0.85\substack{+0.19 \\ -0.16}$ at CMS~\cite{Khachatryan:2014jba} and $1.23\substack{+0.23 \\ -0.20}$ at ATLAS~\cite{ATLAS-CONF-2015-007}, so we take the current bound to be 20\%. These bounds will improve in the future, but ultimately will be systematics limited because of uncertainties in the SM prediction as well as experimental complications. With 3,000 fb$^{-1}$, the expected error on the coupling is 3-5\%~\cite{Dawson:2013bba}, so we take the ultimate expected error on the rate (twice the error on the coupling) to be 10\%. 

\section{EFT Modifications to di-Higgs Production}
\label{sec:EFT}

In this section, we consider the generic effects of new heavy colored particles on di-Higgs production from an EFT perspective. When integrated out, these states will induce the effective operators presented in the introduction in Eq.~(\ref{eq:EFT}). We can then write the relevant couplings contributing to di-Higgs production as
\begin{equation}
 \frac{\alpha_s}{12 \sqrt{2} \pi v} (1 + \kappa_{1}^h+\kappa_2^h+ \dots) h\, G_{\mu\nu}^a G^{\mu\nu a} 
  -  \frac{\alpha_s}{48 \pi v^2}  (1+ \kappa_1^{hh}+ \kappa_2^{hh} + \dots )h^2 \, G_{\mu\nu}^a G^{\mu\nu a}  -\frac{m_h^2}{2 v} h^3.
  \label{eq:EFT-couplings}
\end{equation} 
Here we have defined the relative coupling shifts induced by the higher dimension operators defined in Eq.~(\ref{eq:decouplingOpshhgg}), i.e. $\kappa_1^h = -\kappa_1^{hh} = c_1 (24 \pi/\alpha_s) (v^2 / \Lambda^2)$, $\kappa_2^h = -3 \kappa_2^{hh} = c_2 (48 \pi/\alpha_s) (v^4 / \Lambda^4)$, etc. We would like to understand the extent to which these coupling shifts can modify the di-Higgs production rate while being consistent with the observed SM-like single Higgs production. 

The total di-Higgs production cross section can be written as 
\begin{equation}
\sigma(pp\rightarrow hh) =  \int_{\tau_h}^1  d\tau  \frac{d {\cal L}}{d\tau}  \hat \sigma( \tau s). 
\label{eq:sigtotal}
\end{equation}
Here the gluon parton luminosity is defined as
\begin{equation}
\frac{d {\cal L}}{d\tau}   =  \int_\tau^1 \frac{dx}{x} f_g(x,Q)f_g(\tau/x,Q), \nonumber \\
\label{eq:ggL}
\end{equation}
where $f_g(x,Q)$ is the gluon parton distribution function, with factorization scale $Q$. Throughout this paper we use the MSTW \cite{Martin:2009iq,Martin:2009bu,Martin:2010db} parton distribution functions when calculating the hadronic differential cross sections, with renormalization and factorization scales set to the invariant mass of the di-Higgs system. The partonic cross section in Eq.~(\ref{eq:sigtotal}) is given by
\begin{eqnarray}
\hat \sigma(\hat s)  &   =  &  \frac{\alpha_s^2 \, \hat s \, \beta_h}{2^{15} \, 3^2 \, \pi^3 \, v^4} |A(\hat s)|^2,
\end{eqnarray}
with $\beta_h = (1-4 m_h^2/\hat s)^{1/2}$. With the couplings in Eq.~(\ref{eq:EFT-couplings}), the function $A(\hat s)$ is given by
\begin{equation}
A(\hat s)  =   \frac{ 3 m_h^2 }{ \hat s - m_h^2 }(1+ \kappa_{1}^h+\kappa_2^h+ \dots)  - (1+ \kappa_1^{hh}+ \kappa_2^{hh} + \dots )  . ~~~~~
\label{eq:A}
\end{equation}

\begin{figure}[t]
\centering
\includegraphics[height=2.8in]{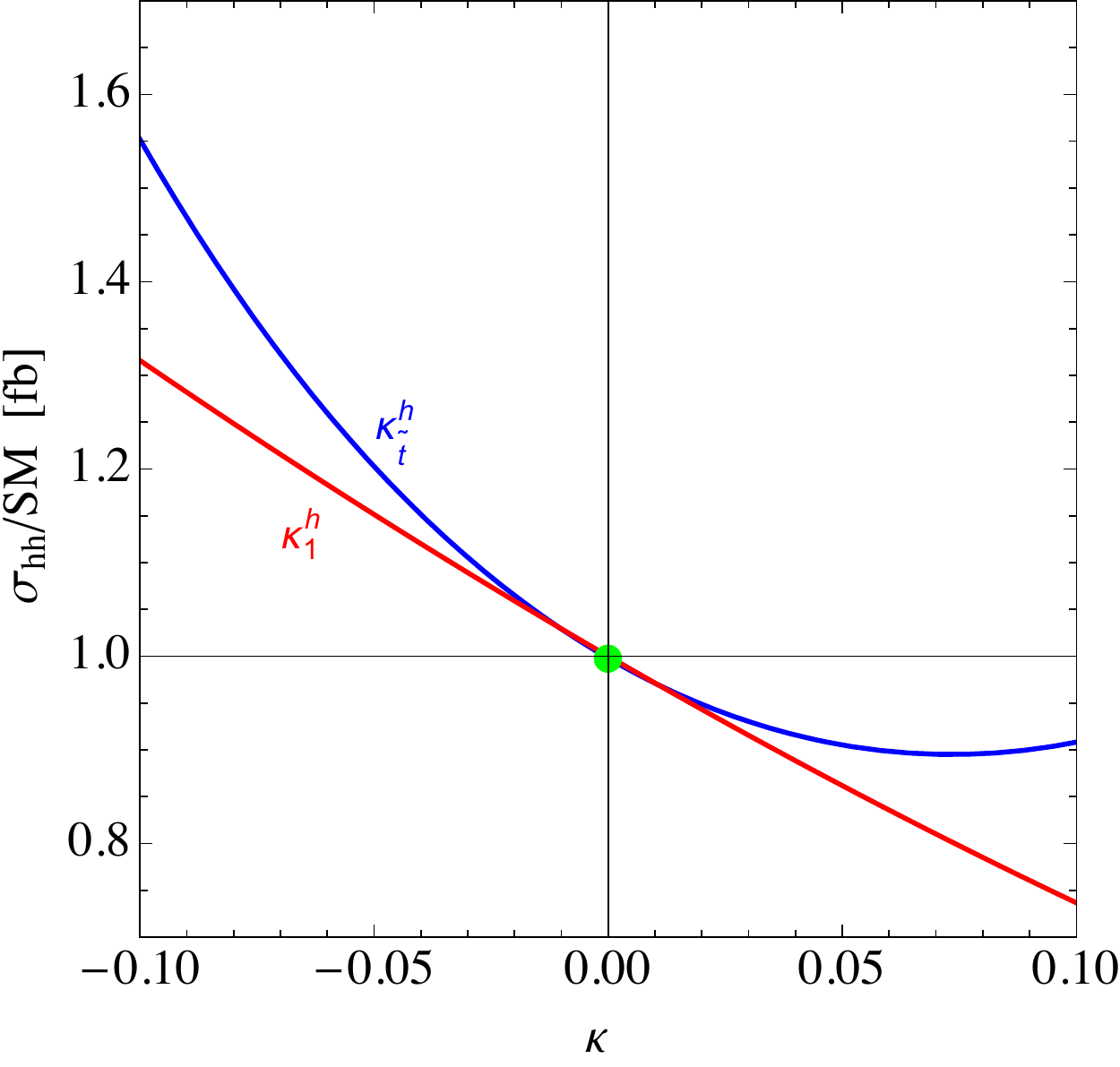} 
\caption{Di-Higgs production cross section relative to the SM value as a function of $hGG$ coupling deviation in an EFT dominated by the leading dimension six operator (red, $\kappa_1^h$) computed in Sec.~\ref{sec:EFT}, and for heavy stops using the low energy theorem (blue, $\kappa_{\tilde t}^h$) computed in Sec.~\ref{sec:GenStopMod}. 
The coupling deviation $\kappa$ is taken in range $[-0.1,0.1]$ as suggested by the LHC Run 1 Higgs data. 
}
\label{fig:LET}
\end{figure}

Let us consider the case in which the new heavy colored states decouple from the Higgs as their mass is raised. This will happen if these states primarily obtain their mass from sources other than electroweak symmetry breaking. In this case, there is a separation of scales, $v \ll  \Lambda$, and the EFT expansion in Eq.~(\ref{eq:EFT}) is a useful one. The leading dimension 6 operator dominates over the dimension 8 (and higher) operators, $c_1 v^2/\Lambda^2\gg c_2 v^4 /\Lambda^4$ and there is a well-defined relation between the single and double Higgs production rate via gluon fusion in terms of the parameter $\kappa_{1}^h$, which is $\kappa_1^h = -\kappa_1^{hh}$. In Fig.~\ref{fig:LET} we plot the ratio of the di-Higgs production cross section to the SM prediction as a function of the $hGG$ coupling shift $\kappa_{1}^h$ arising in the EFT. As the Run 1 Higgs results restricts $|\kappa_1^h| <10\%$, we observe that an enhancement or suppression of the di-Higgs production rate of order 30$\%$ is still allowed by the data within the context of the EFT. In this case, one can easily understand the origin of the enhancement (suppression) when $\kappa_{1}^h$ is negative (positive) by examining the interference between the box and triangle diagrams 
(see Fig.~\ref{fig:dihiggsEFTdiagrams}) via the function $A(\hat s)$ in Eq.~(\ref{eq:A}). For instance, when $\kappa_1^h$ is negative, the smaller triangle amplitude is suppressed, while $\kappa_1^{hh} = -\kappa_1^h$ is positive and the dominant box amplitude is enhanced. This implies that the interference between the amplitudes is reduced in comparison to the SM and the di-Higgs rate is enhanced. 

There are other qualitatively distinct cases to consider. The first is when the new heavy colored states do not decouple from the Higgs as their mass is raised. This will occur if the new states obtain a substantial portion of their mass from electroweak symmetry breaking. In the language of the EFT, each operator in Eq.~(\ref{eq:EFT}) is of similar size and thus the expansion is not useful from a practical point of view. This type of non-decoupling behavior is of course very familiar from the top quark contribution to the $hGG$ and $hhGG$ couplings. In this case it is instead necessary to specify the model for the new heavy colored states and apply the low energy theorems~\cite{Shifman:1979eb,Kniehl:1995tn,Gillioz:2012se,Kribs:2012kz}, as seen for light stops in Sec.~\ref{sec:GenStopMod}.

Finally, the last case to consider is when the new states are light such that neither the EFT nor LET descriptions are valid. In the case of di-Higgs production, this occurs when the masses of the new states in the loop are similar to the characteristic invariant mass of the di-Higgs system under consideration. In this situation it is necessary to specify the model under consideration and compute the full one loop contribution to di-Higgs production. This is carried out for light stops in Sec.~\ref{sec:loop}.

\section{Heavy Stop Modifications: Low Energy Theorem}
\label{sec:GenStopMod}

For the remainder of the paper we specialize to the case of stops in supersymmetry, which provides a well-motivated, concrete example of new colored particles with significant couplings to the Higgs.
As is well known, the MSSM requires two Higgs doublets. However, motivated by the lack of evidence for new scalars and the fact that the Higgs production and decay rates are measured to be near the SM value, we take the 125 GeV Higgs to be the lightest neutral scalar boson and work in the decoupling limit. For the light stops that we consider in this work, we will typically not be able to obtain the 125 GeV Higgs mass in the MSSM. 
However, there are many possible scenarios that raise the Higgs mass including, for example, the NMSSM (for a review see~\cite{Maniatis:2009re,Ellwanger:2009dp}) or non-decoupling $D$-terms~\cite{Batra:2003nj,Maloney:2004rc}. 
Therefore, we take the Higgs potential, particularly the triple Higgs coupling, to be that of the Standard Model in order to focus on the stop contributions.

We begin by describing our conventions for the stop sector. The stop mass matrix is given by
\begin{equation}
{\cal M}_{\tilde t}^2 = 
\left(  
\begin{array}{cc}
m_{LL}^2 & m_{LR}^2 \\
m_{LR}^2 & m_{RR}^2
\end{array}
    \right),\label{eq:stopmassmatrix}
\end{equation}  
where we have defined
\begin{eqnarray}
m_{LL}^2 & = &  m_{Q_3}^2 + y_t^2 v_u^2 +\tilde \Delta_Q (v_d^2-v_u^2), \nonumber \\
m_{RR}^2 & = &  m_{U_3}^2 + y_t^2 v_u^2 +\tilde \Delta_U (v_d^2-v_u^2), \nonumber \\
m_{LR}^2 & = &  y_t (A_t v_u -\mu v_d )\equiv m_tX_t,   
\end{eqnarray}
with $\tilde \Delta_Q = \tfrac{1}{2}(\tfrac{1}{2} g^2 - \tfrac{1}{6} g'^2)$, $\tilde \Delta_U = \tfrac{1}{2}( \tfrac{2}{3} g'^2)$. We also take $\sqrt{v_u^2+v_d^2}=v=174$ GeV and define $\tan\beta\equiv v_u/v_d$. This matrix can be diagonalized, with eigenvalues $m_1$ and $m_2$ satisfying $m_2>m_1$, by performing a rotation of the basis by the angle $\theta$ defined by
\begin{equation}
\begin{array}{cc}
\displaystyle \cos 2\theta=\frac{m_{LL}^2-m_{RR}^2}{m_2^2-m_1^2},~~ &~~  \displaystyle \sin 2\theta=-\frac{2m_tX_t}{m_2^2-m_1^2}
\end{array}.
\end{equation}

In this section we examine the generic corrections to the di-Higgs production rate in the limit that the stops are heavy in comparison to the typical di-Higgs invariant mass. As alluded to in the previous section, the stops can in general exhibit non-decoupling behavior as their masses are raised if the $X_t$ parameter is also raised in a correlated fashion. This is analogous to the case of the top quark in the SM. Because of this potential non-decoupling behavior, we we apply the Low Energy Theorem (LET)~\cite{Shifman:1979eb,Kniehl:1995tn,Gillioz:2012se,Kribs:2012kz} to derive the couplings of the Higgs to gluons induced by stops. The starting point is the stop threshold contribution to the running of $\alpha_s$. After canonical normalization of the gluon field, we obtain the following effective Lagrangian:
\begin{eqnarray}
{\cal L}&  \supset &
 \frac{\alpha_s b^c_0 }{16 \pi} \, \left[ \log \det {\cal M}_{\tilde t}^2 \right] \, G_{\mu \nu} G^{\mu\nu},
\label{eq:Lgg}
\end{eqnarray}
where $b_0^c = \tfrac{1}{6}$ is the QCD beta function coefficient for stops. 

Using Eq.~(\ref{eq:Lgg}) we determine the couplings of the Higgs $h$ to gluons generated from stops by Taylor expanding around $v_u$ and $v_d$ in the Higgs fluctuations. Including the dominant SM top quark contribution, we arrive at the following effective Lagrangian describing the Higgs couplings to gluons:
\begin{equation}
{\cal L} = \frac{\alpha_s}{12 \sqrt{2} \pi v} (\kappa_{t}^{h} + \kappa_{\tilde t}^{h} ) h\, G_{\mu\nu} G^{\mu\nu} 
  -  \frac{\alpha_s}{48 \pi v^2}  (\kappa_{t}^{hh} + \kappa_{\tilde t}^{hh} )h^2 \, G_{\mu\nu} G^{\mu\nu}.    
\end{equation}
The coefficients $\kappa_{t}^{h}$, $\kappa_{t}^{hh}$ ($\kappa_{\tilde t}^{h}$, $\kappa_{\tilde t}^{hh}$) encode the top quark (stop) contributions to the $hGG$ and $h^2GG$ couplings. In particular, for the stop contribution we have 
\begin{eqnarray}
\kappa_{\tilde t}^h & \equiv &  \frac{v}{8}  \left( c_\alpha \frac{\partial}{\partial v_u} - s_\alpha \frac{\partial}{\partial v_d}  \right) \log \det {\cal M}_{\tilde t}^2, 
\qquad\\
\kappa_{\tilde t}^{hh} & \equiv &- \frac{v^2}{8} \left( c^2_\alpha \frac{\partial^2}{\partial v_u^2}  +  s^2_\alpha \frac{\partial^2}{\partial v_d^2} 
- 2 s_\alpha c_\alpha \frac{\partial^2}{\partial v_u \partial v_d}   \right) \log \det {\cal M}_{\tilde t}^2.  \nonumber 
\label{eq:xi}
\end{eqnarray}
Here $\alpha$ is the mixing angle between the light and heavy CP-even Higgs bosons.
Neglecting the small contributions from $D$-terms ($g,g'\rightarrow 0$) and taking the decoupling limit ($\alpha \rightarrow \beta -\pi/2$) we obtain 
\begin{eqnarray}
\kappa_{\tilde t}^{h}& = & \frac{1}{4} \frac{ m_t^2 \left( m_{1}^2 + m_{2}^2 - X_t^2 \right)  }{m_{1}^2 m_{2}^2} ,  \\
\kappa_{\tilde t}^{hh} & = & -\frac{ m_t^4}{m_{1}^2 m_{2}^2 }
 \bigg\{ 1
 + \frac{\left(m_{1}^2 +m_{2}^2-X_t^2\right)}{ 4m_t^2 }
 - \frac{\left(m_{1}^2+m_{2}^2-X_t^2\right)^2}{2 m_{1}^2 m_{2}^2 } 
 \bigg\} \nonumber \\
 & = &  \kappa_{\tilde t}^{h}(8 \, \kappa_{\tilde t}^{h}-1) -\frac{ m_t^4}{m_{1}^2 m_{2}^2 }, 
 \label{eq:c}  
\end{eqnarray}
where in the final step we have written $\kappa_{\tilde t}^{hh} $ in terms of $\kappa_{\tilde t}^{h}$. 
These stop-induced contributions are to be compared to the top quark contributions, which in the decoupling limit are $\kappa_{t}^{h} = \kappa_{t}^{hh} = 1$. Therefore, the parameters $\kappa_{\tilde t}^{h}$ and $\kappa_{\tilde t}^{hh} $ measure the relative coupling shift from the SM values in an analogous way to the EFT coupling shifts defined in the previous section. 
We see from the last line in Eq.~(\ref{eq:c}) that a definite correlation exists between the $hhGG$ and the $hGG$ couplings, and in the limit of heavy stops, $m_{1,2} \gg m_t$, the $hhGG$ coupling shift is fully determined by $\kappa_{\tilde t}^{h}$. 

As emphasized above, the current Run 1 data probe deviations in the $hGG$ coupling at the 10$\%$ level, i.e., 
$|\kappa_{\tilde t}^{h}| \lesssim 10 \%$. One can use this constraint to estimate the allowed size of the corrections to the di-Higgs rate from heavy stops by using Eq.~(\ref{eq:c}). This is shown in Fig.~\ref{fig:LET}, where we observe that ${\cal O}(50\%)$ corrections are possible when the $hGG$ coupling is smaller than its SM value by about 10$\%$. The behavior can be easily understood by examining the couplings $\kappa_{\tilde t}^{h} $ and $\kappa_{\tilde t}^{hh} $ and accounting for the interference between the two diagrams depicted in Fig.~\ref{fig:dihiggsEFTdiagrams}. For instance, when $\kappa_{\tilde t}^h$ is negative the $s$-channel Higgs exchange amplitude is slightly suppressed compared to its SM value, while the larger-in-magnitude contact diagram is instead mildly enhanced (since $\kappa_{\tilde t}^{hh}$ is positive when $\kappa_{\tilde t}^h$ is negative, assuming the stops are heavy). Therefore, the interference between the diagrams is less effective leading to the enhanced rate in this region, as shown in Fig.~\ref{fig:LET}.

In Fig.~\ref{fig:LET} we can also see the importance of the non-decoupling behavior by comparing the EFT to the LET calculation. Because $A$-terms can cause the stops to get a large fraction of their mass from electroweak symmetry breaking even if they are relatively heavy, different and potentially larger effects in di-Higgs can be induced. Therefore, if a deviation is observed but no on-shell states are discovered, the size of the deviation could disentangle different types of decoupling vs non-decoupling new physics scenarios.

\section{Light Stop Modifications: Full Loop Calculation}
\label{sec:loop}

\begin{figure}[t]
\centering
\includegraphics[height=2.8in]{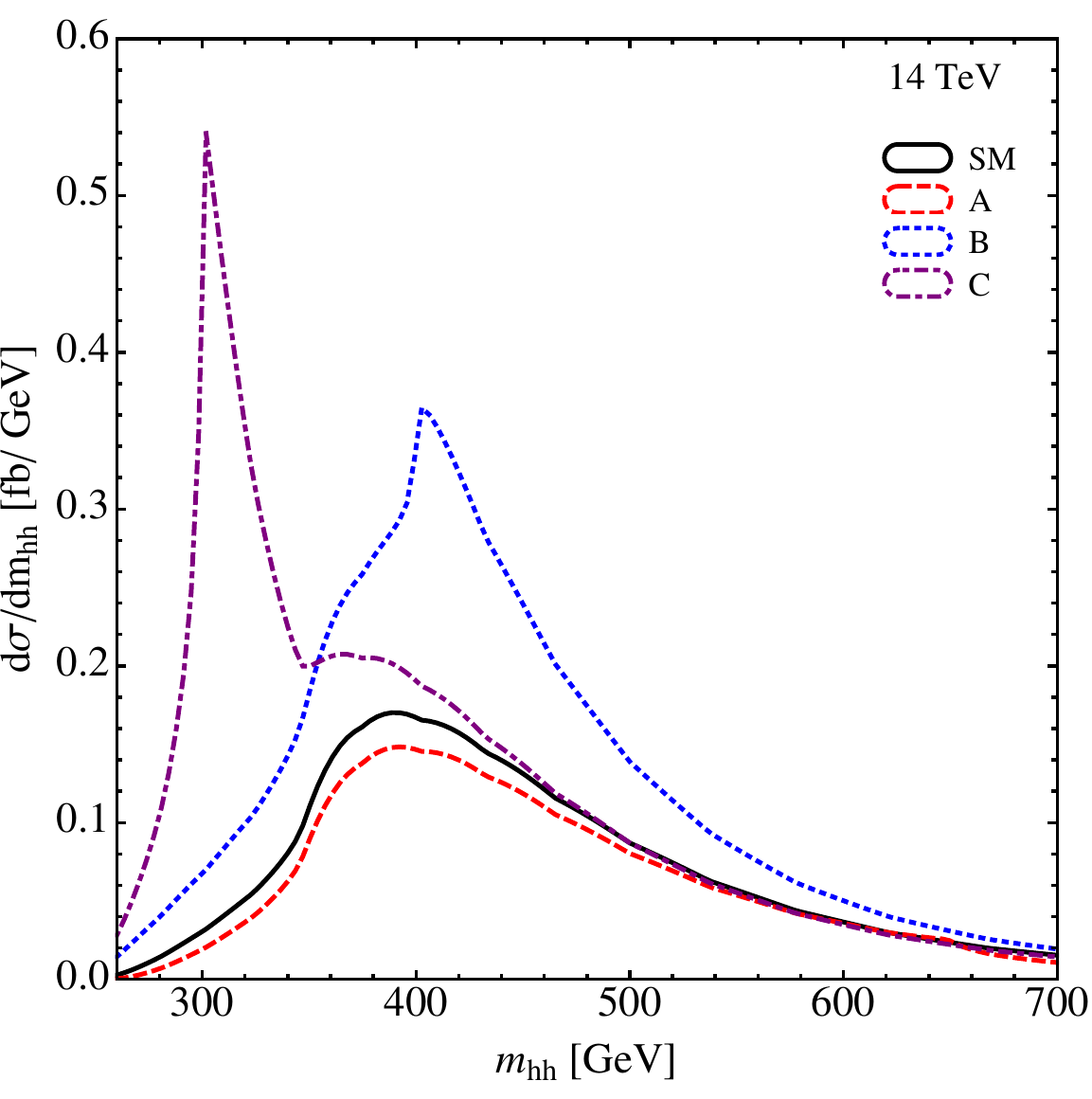} 
\caption{Invariant mass spectrum for di-Higgs events at the LHC14. We show spectra for the SM, and the benchmark points: A) Both stops near the weak scale and current constraints satisfied, $m_1 = 325$ GeV, $m_2 = 500$ GeV, $\sin \theta = 0.4$, B) One stop heavy, current constraints satisfied and a large enhancement of di-Higgs production through tuning of the mixing angle, $m_1 = 200$ GeV, $m_2 = 1000$ GeV, $\sin \theta = 0.223$, C) One stop light and single Higgs production constraints \textit{not} satisfied, $m_1 = 150$ GeV, $m_2 = 1000$ GeV, $\sin \theta = 0$.}
\label{fig:Spectrum}
\end{figure}

Finally, we consider the effects of light stops on the di-Higgs rate, which requires a full one loop analysis. To calculate the parton-level single Higgs and di-Higgs production cross sections we implemented the SM+Stops model described above into the {\sc FeynArts} package~\cite{Hahn:2000kx,Hahn:1998yk} and employed the {\sc{FeynArts}}, {\sc{FormCalc}}, and {\sc{LoopTools}} suite of packages~\cite{Hahn:2000kx,Hahn:1998yk} to calculate the amplitudes and evaluate loop functions. We used the MSTW \cite{Martin:2009iq,Martin:2009bu,Martin:2010db} parton distribution functions when calculating the hadronic differential cross sections, with renormalization and factorization scales set to the invariant mass of the di-Higgs system. For the spectra in \Fig{fig:Spectrum} we use constant K-factors to normalize our LO result to the NLO results in 
\cite{Baglio:2012np}. However, these K-factors cancel out in all other plots as only ratios of the BSM rate with the SM rate are shown.
We have also cross checked our results using the full one-loop MSSM computations of Refs.~\cite{Belyaev:1999mx,BarrientosBendezu:2001di}, finding good agreement.\footnote{We differ in the writing of the function $F_3$ defined in equation (B.2) of \cite{BarrientosBendezu:2001di}:
\begin{align}
F_3(s,t,h_1,h_2,m^2_{\tilde{q}_i},m^2_{\tilde{q}_j})=& -s(t+m^2_{\tilde{q}_i}) C^{00}_{iii}(s)+sm^2_{\tilde{q}_i}C^{00}_{jjj}(s)-tt_1C^{h_10}_{ijj}(t) -tt_2C^{h_20}_{ijj}(t)\nonumber\\
&+(t^2-h_1h_2)C^{h_1h_2}_{iji}(s) -2stm^2_{\tilde{q}_i}D^{h_1h_200}_{jijj}(s,t) \nonumber\\
&+\left[st^2-2t_1t_2m^2_{\tilde{q}_i}+s(m^2_{\tilde{q}_i}-m^2_{\tilde{q}_j})^2 \right] D^{h_1h_200}_{ijii}(s,t)\nonumber\\
&+\frac{s}{2}\left[ p_T^2(m^2_{\tilde{q}_i}+m^2_{\tilde{q}_j})+ (m^2_{\tilde{q}_i}-m^2_{\tilde{q}_j})^2\right] D^{h_10h_20}_{ijji}(t,u)+ (t\leftrightarrow u).
\end{align}
}  

We begin by examining some benchmark models and their effect on the di-Higgs invariant mass spectra. In the SM, the amplitude for di-Higgs production vanishes at threshold because of a cancellation between the top box diagram and a triangle diagram that utilizes the triple Higgs coupling~\cite{Li:2013rra,Dicus:2015yva,Dawson:2015oha}, and this is true for any field content as long all masses are acquired via the Higgs vacuum expectation value. Therefore, the invariant mass distribution in the SM is very small near threshold and grows to a peak near $m_{hh} \sim 2 m_t$, as we see in \Fig{fig:Spectrum}. Generic new physics that mediates one-loop di-Higgs production will spoil this cancellation, so light colored particles can lead to large deviations near threshold. We demonstrate this for some benchmark cases in \Fig{fig:Spectrum}.

Benchmark A has $m_1 = 325$ GeV, $m_2 = 500$ GeV, $\sin \theta = 0.4$: it has both stops light but the mixing angle is such that the the rate of $gg\rightarrow h$ is only enhanced by $\sim15\%$ and the $h\rightarrow\gamma\gamma$ rate is within 5\% of the SM value. This is a typical case where even having light stops the di-Higgs spectrum looks SM-like, and the total rate is $\sim 86\%$ of the SM; a modification unobservable at the LHC. This also illustrates the effect found in Fig.~\ref{fig:LET} that the sign of the modification in single production is anti-correlated with that of the di-Higgs rate. Benchmark B has one light stop and one heavy stop, $m_1 = 200$ GeV, $m_2 = 1000$ GeV, $\sin \theta = 0.223$, with the mixing angle carefully tuned to give a large enhancement in the di-Higgs rate while still being allowed by single Higgs data. The largest enhancement in the spectrum occurs around 400 GeV where the lighter stop in the loop can go on-shell. The total di-Higgs rate is enhanced by $\sim 70\%$, the single Higgs rate is reduced by $\sim 20\%$, and the di-photon modification is small. 

In benchmark C we show the generic but excluded case with one light stop: $m_1 = 150$ GeV, $m_2 = 1000$ GeV, $\sin \theta = 0$. Here the cancellation in the matrix element at threshold discussed in the introduction is spoiled and there is a large cross section enhancement at low invariant mass. The total cross section is enhanced by $\sim 90\%$, but the single Higgs rate is also enhanced by $\sim 80\%$. 

We now discuss the expected modifications to the di-Higgs production rate as a function of more general stop sector parameters. Throughout we consider corrections to single Higgs and di-Higgs production. We will also consider two bins of di-Higgs invariant mass: $260 < m_{hh} < 350$ GeV and $260 < m_{hh} < 2000$ GeV. The first region is motivated because for light stops, the di-Higgs invariant mass spectrum can deviate significantly from the SM prediction for $m_{hh} < 2 m_t$, as was illustrated in \Fig{fig:Spectrum}. Thus, although the total number of signal events may be smaller, when constraining new non-resonant contributions to di-Higgs production it may help to focus on di-Higgs invariant mass bins close to the threshold for production as this is where corrections are likely to be greatest. We also consider the full invariant mass regime to make contact with previous phenomenological studies that also do so. 

In this section we consider corrections only at $14$ TeV and provide contours for $100$ TeV in \App{sec:100TeV}. The total di-Higgs production cross section increases substantially when going from $14$ to $100$ TeV, which is essentially due to the increased gluon luminosity. This is the main reason that sensitivity to di-Higgs production improves significantly with a $100$ TeV proton-proton collider when compared to the LHC. However, for light stops the \emph{ratio} of cross section modifications to the SM cross section remains roughly the same for both colliders. The reason for this is that although the total gluon luminosity in both cases is significantly different, the gradient of the gluon luminosity with respect to parton center of mass energy is not significantly different in the region of interest for di-Higgs production. Thus, when integrating over the parton distribution functions the increased gluon luminosity is roughly a constant factor, especially in the low invariant mass bin, and hence when the ratio of total cross section with stops to the total cross section in the SM is taken this factor essentially drops out. Therefore, the fractional corrections are very similar at $14$ and $100$ TeV. This does not persist whenever the stops are heavy and features in the invariant mass distribution appear at large $m_{hh}$ where the gluon luminosity between $14$ and $100$ TeV is significantly different, but in this case the corrections are typically smaller than the expected sensitivity. Thus, for the fractional corrections to the total cross section the $14$ TeV results are also roughly illustrative of the $100$ TeV result, although the expected sensitivity is increased at higher center of mass energy, so it should be kept in mind that contours of different di-Higgs cross section are appropriate in this case.

\begin{figure}[h]
\centering
\includegraphics[height=2.6in]{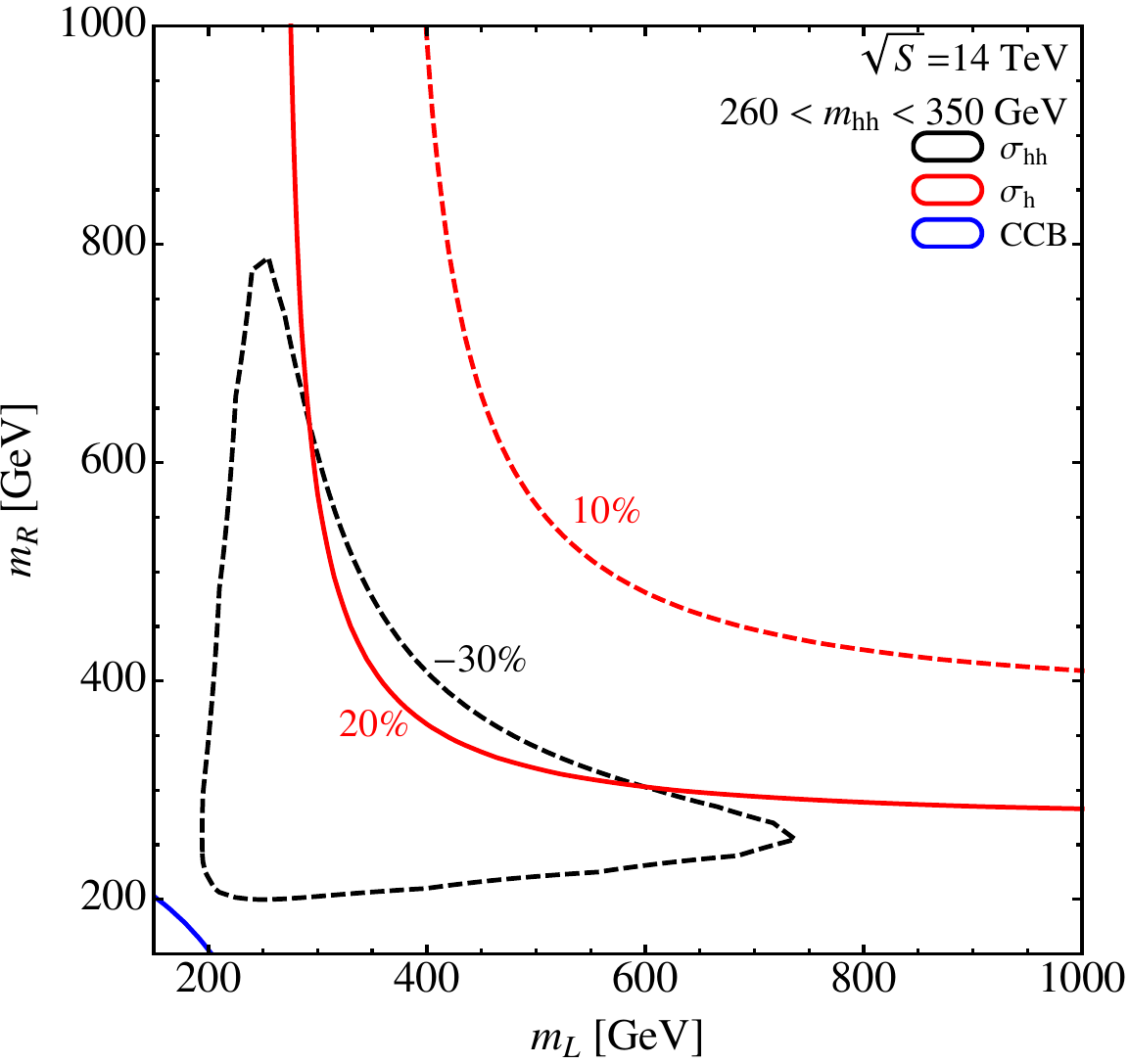} \qquad \includegraphics[height=2.6in]{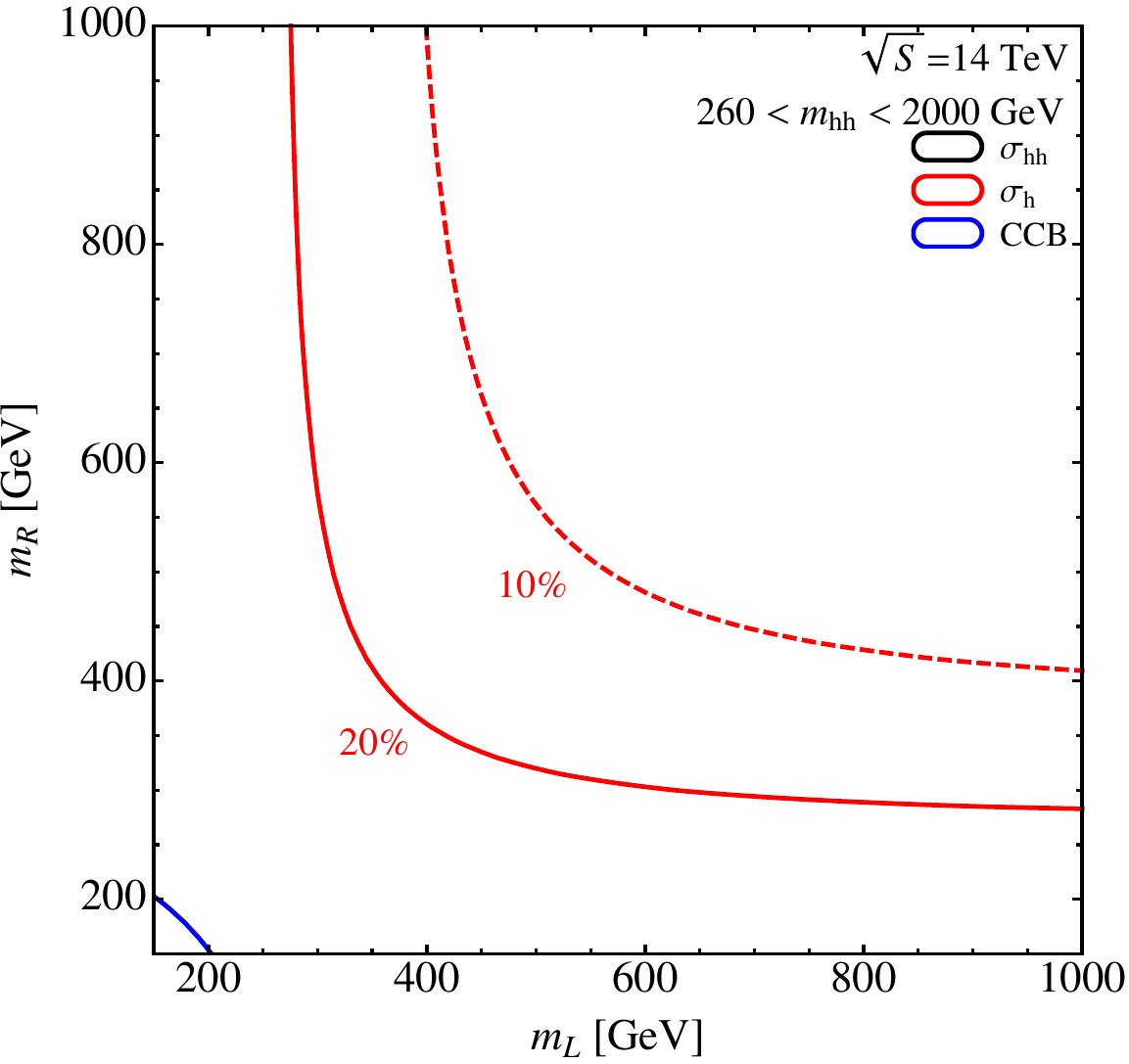}
\caption{Percentage corrections to the single Higgs (red) and di-Higgs (black) production cross sections at $\sqrt{s} = 14$ TeV in a low energy bin with invariant masses $260 < m_{hh} < 350$ GeV (left) and a wide bin with $260 < m_{hh} < 2000$ GeV (right). For the wide energy bin the corrections fall below the benchmark sensitivity for all soft masses shown. Both stop soft masses are varied independently and the $A$-terms are set to zero. The masses on the axes are the physical masses of the left- and right-handed stops. Small differences between left and right-handed stops due to different $D$-term couplings can be seen. We also show blue contours of the approximate color-breaking vacuum constraint described in \Sec{sec:CCB}.}
\label{fig:m1m2}
\end{figure}

\begin{figure}[h]
\centering
\includegraphics[height=2.6in]{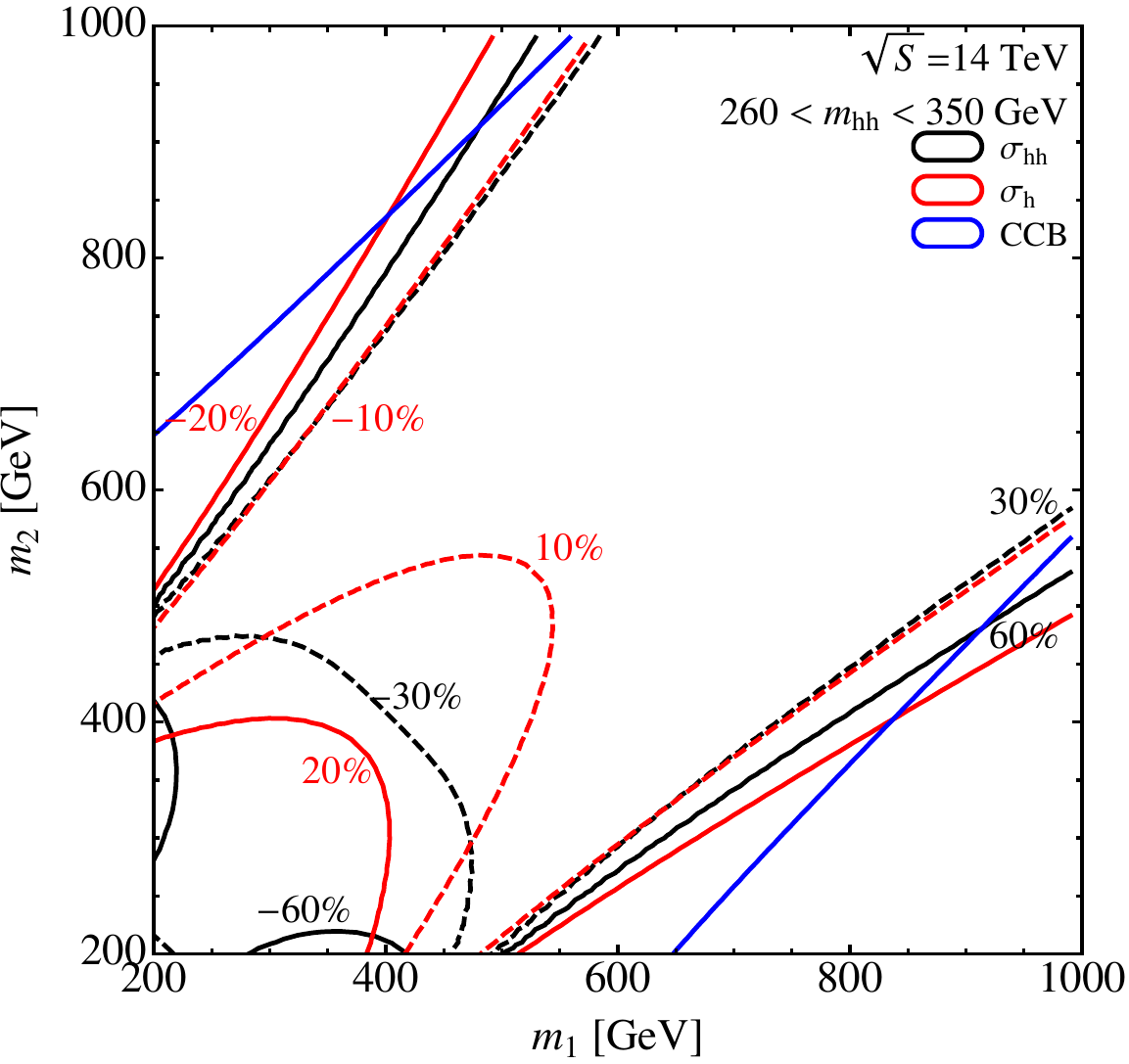} \qquad \includegraphics[height=2.6in]{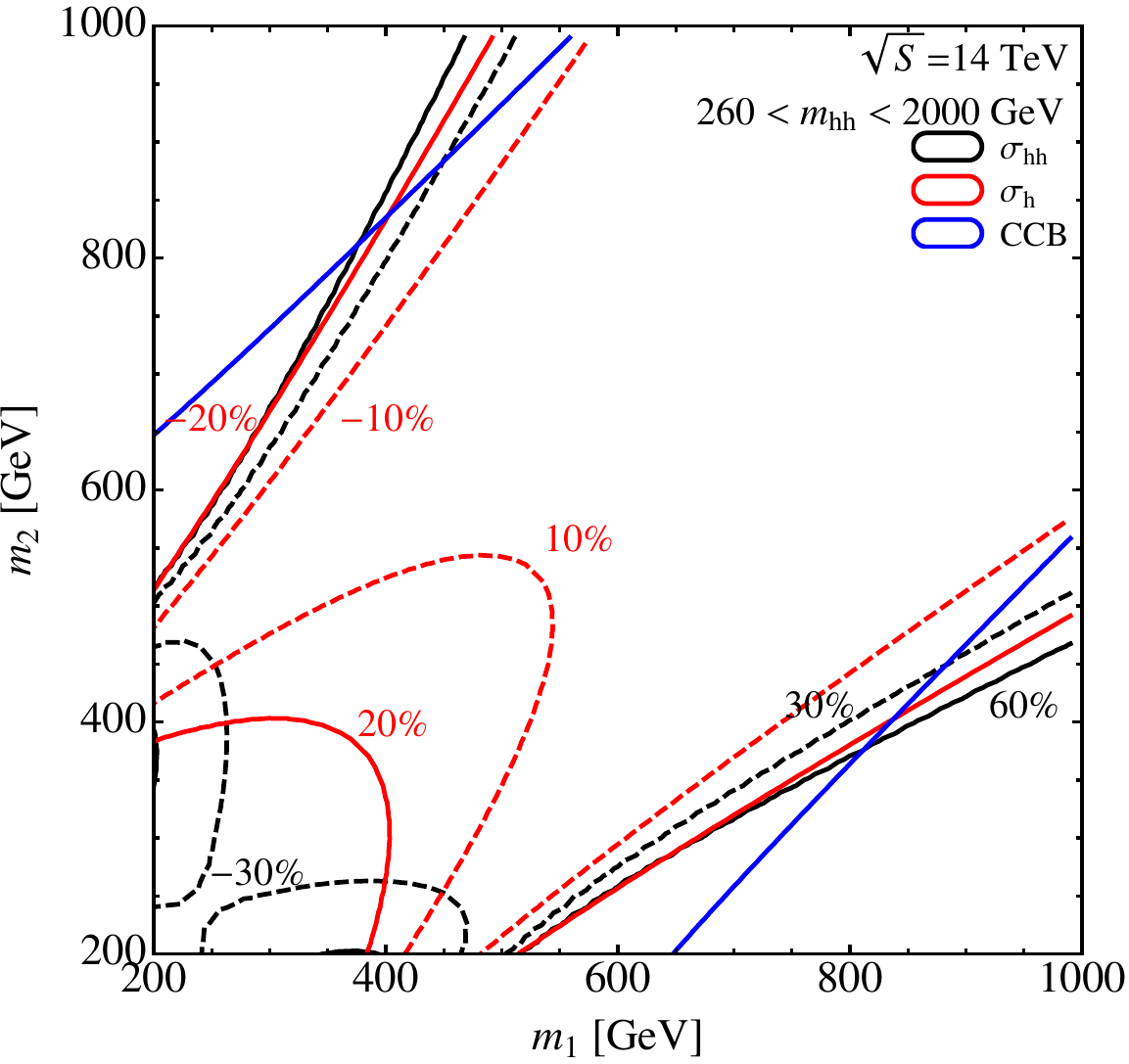}
\caption{As in \Fig{fig:m1m2} with the exception that both stop soft masses are set equal and the $A$-terms are varied. In both cases regions which lead to a $\sim -20 \%$ change in the single production rate typically imply a $\sim 30 \%$ change in the pair production rate. The approximate color breaking vacuum constraint shown in blue is relevant for large mass splittings due to the large $X_t$-terms.}
\label{fig:mA}
\end{figure}

\begin{figure}[t]
\centering
\includegraphics[height=2.6in]{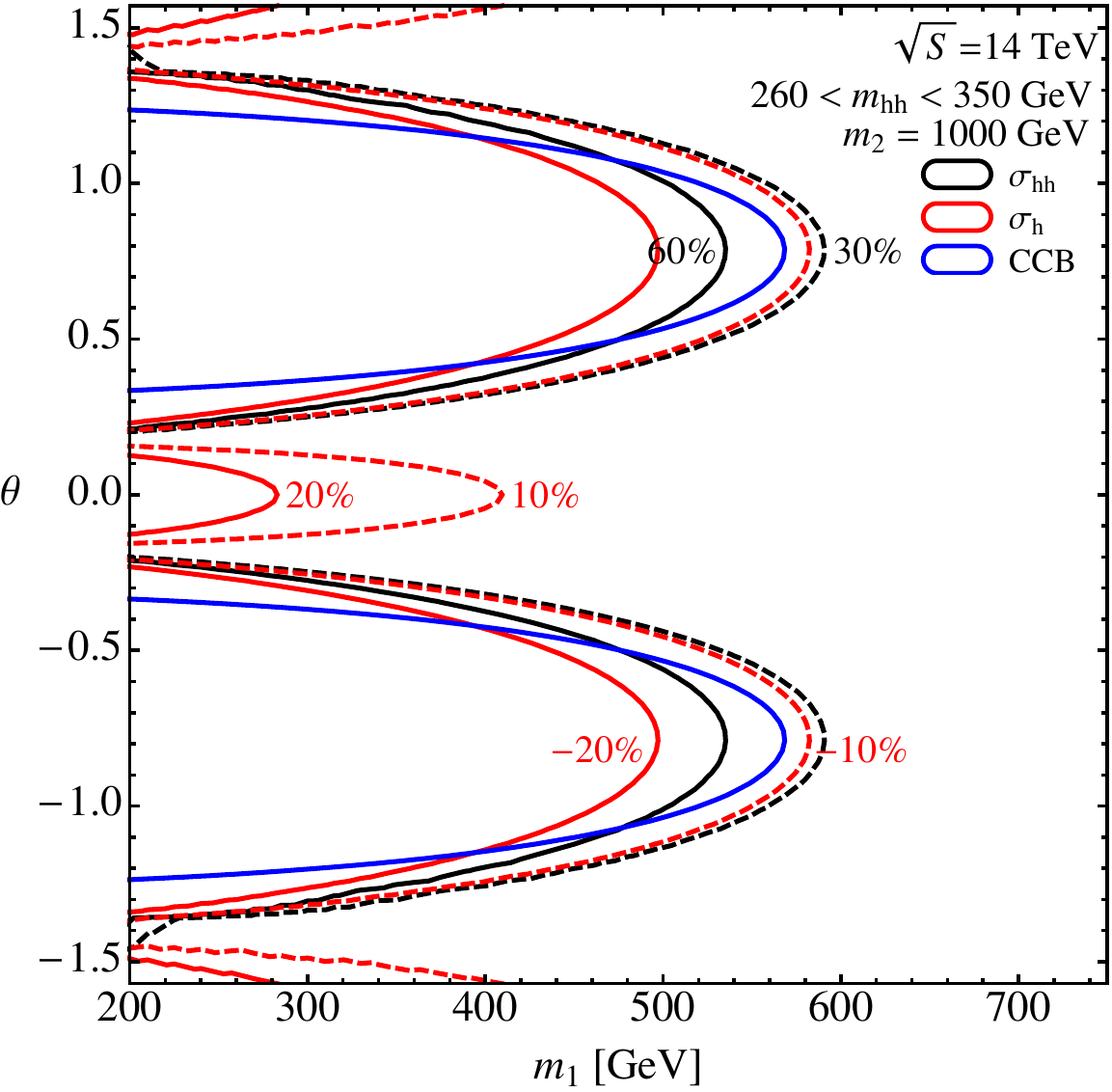} \qquad \includegraphics[height=2.6in]{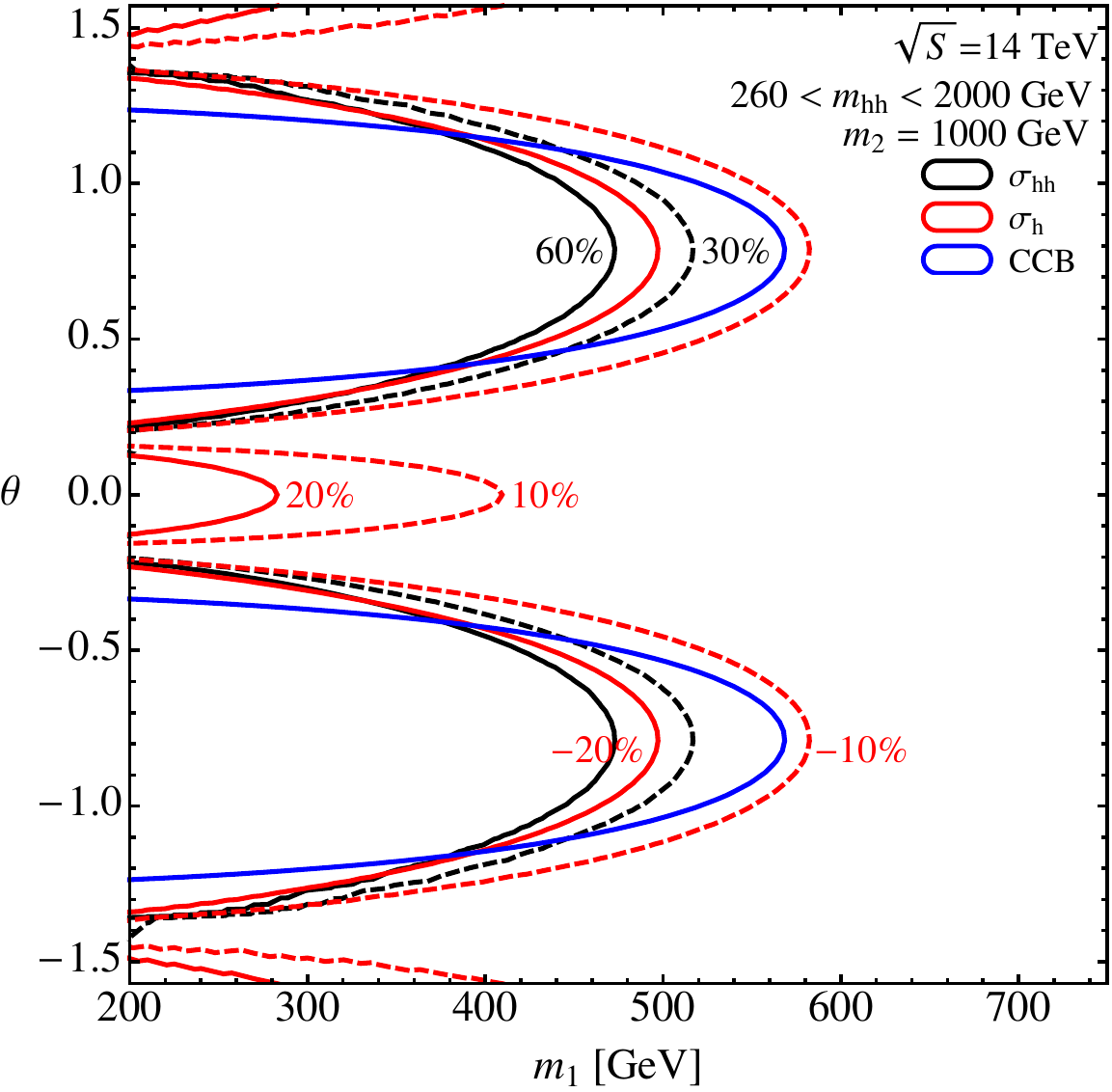}  \\ 
\includegraphics[height=2.6in]{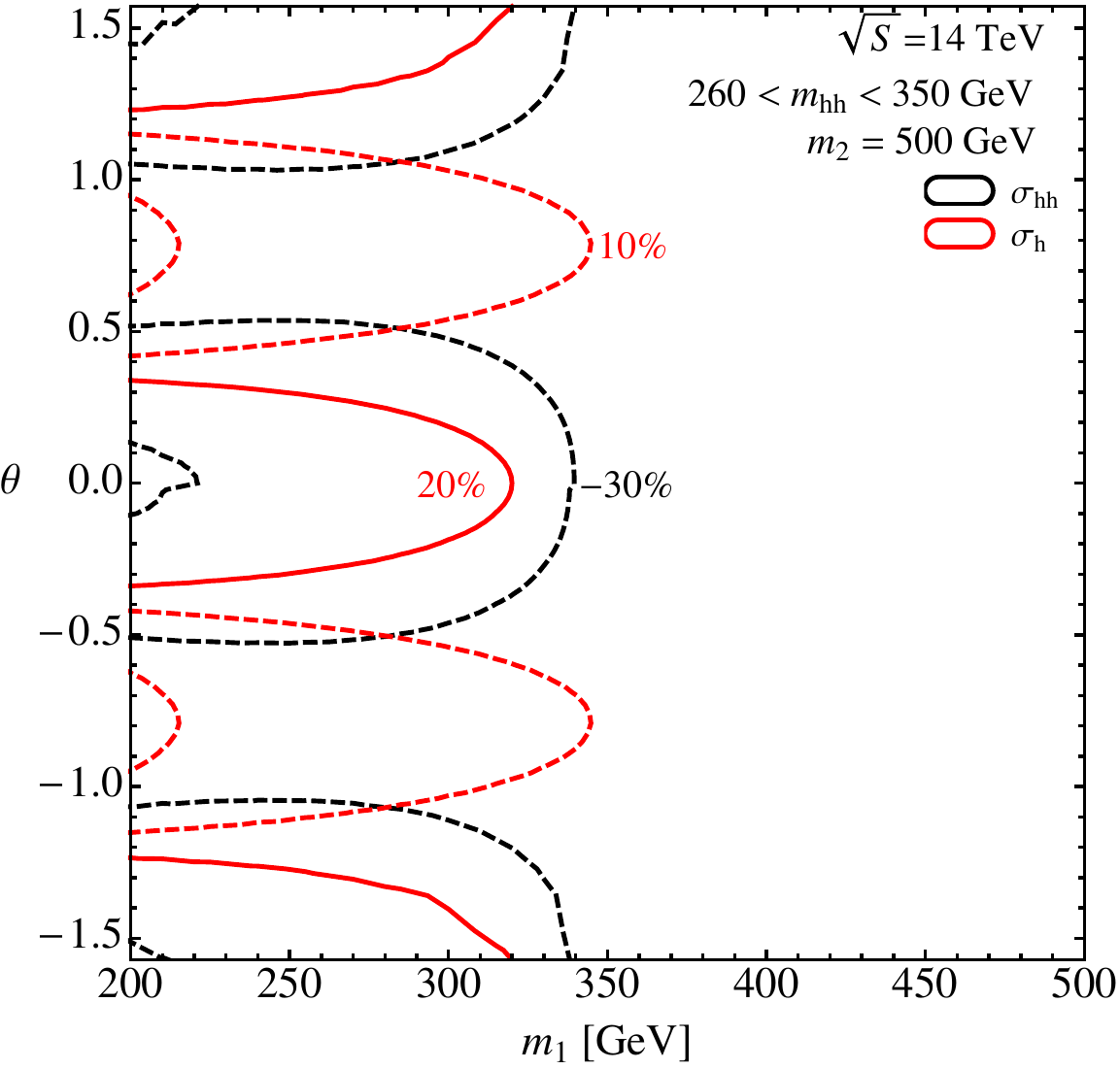} \qquad \includegraphics[height=2.6in]{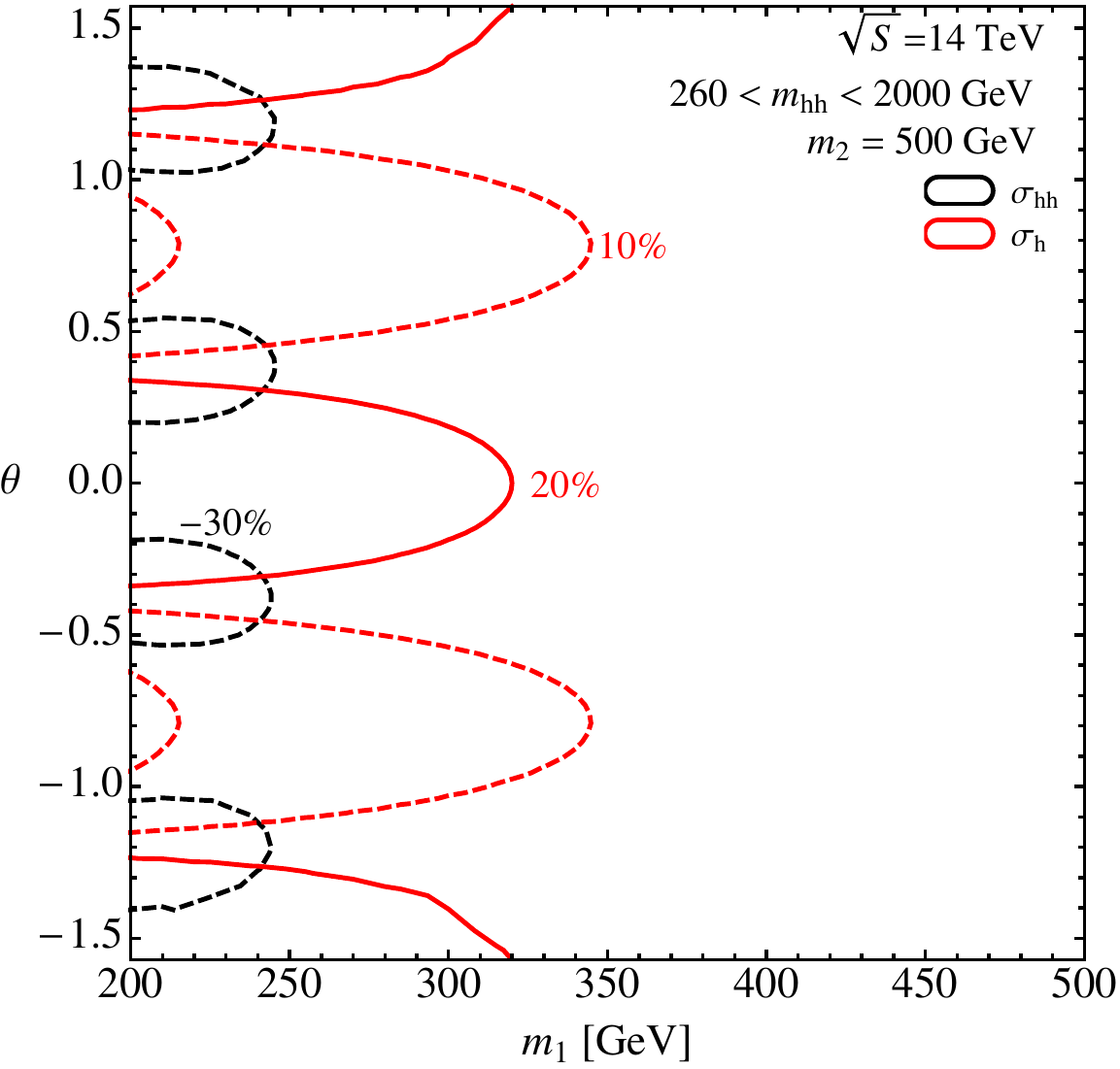} 
\caption{As in \Fig{fig:m1m2} with the exception that the heavy stop mass is fixed at $1000$ GeV (upper panels) and $500$ GeV (lower panels) and the light stop mass and mixing angle are varied.}
\label{fig:mST}
\end{figure}

In general the stop parameter space can be described by three physical parameters, such as the two stop mass eigenvalues $m_1,m_2$, and the mixing angle, or alternatively the two soft masses $\widetilde{m}_L,\widetilde{m}_R$ and the mixing parameter $X_t$. To plot the corrections a projection down to a two-dimensional subspace is necessary. Results for a variety of projections for the full loop calculation are shown in \Fig{fig:m1m2}, \Fig{fig:mA}, and \Fig{fig:mST}. In \Fig{fig:m1m2} the stop mixing $X_t$-terms are set to zero and only the physical mass eigenvalues are varied. In \Fig{fig:mA} the two soft masses are set equal, $\widetilde{m}_L = \widetilde{m}_R$, and varied and the $X_t$-term is also varied. The results are shown in the basis of physical masses. In \Fig{fig:mST} we fix the mass eigenvalue of the heavy stop to a benchmark value and then vary the light stop mass and the stop mixing angle.

In all of the figures a consistent picture emerges. Cases which lead to an observable deviation in the di-Higgs production rate also typically have observable deviations in the single Higgs production rate. Furthermore, in the `blind spot' region where the single Higgs corrections are small the di-Higgs corrections are also typically suppressed unless both stops are quite light, which is consistent with our understanding based on the EFT arguments in \Sec{sec:intro}. Thus, in order to indirectly constrain the existence of light stops which may have evaded direct detection at the LHC, the single Higgs production and di-Higgs production processes are highly complementary indirect probes and the strongest indirect constraints would arise from the combination of the two. Furthermore, taking into account current constraints, \Fig{fig:m1m2}, \Fig{fig:mA}, and \Fig{fig:mST} suggest that tuned regions of parameter space may remain after LHC8 in which observable non-resonant contributions to di-Higgs deviations may still arise at LHC14 from stops.

It is also interesting that, as advertised previously, in \Fig{fig:mA} and \Fig{fig:mST} it is clear that deviations relative to the SM may be significantly larger in low invariant mass bins than they are for the total cross section. However, due to the smaller signal rate, the statistics will be lower in the low mass bin than for the total cross section. Thus in a collider analysis aimed at indirectly constraining stop squarks it may be necessary to study a number of invariant mass cuts to determine the optimal constraint.

Finally, in \Fig{fig:mST} it is clear that if both stops are light the standard `blind spot' in stop contributions to single Higgs productions may be closed by constraining di-Higgs production. This is consistent with our EFT discussion in \Sec{sec:intro} as even when the stop loop contributions to the $hGG$ coupling have been tuned to precisely zero there will remain contributions to the $h^2GG$ coupling coming from a dimension-8 operator. Thus the $h^2GG$ coupling in the blind spot will typically be $\mathcal{O} (m_t^4/{m_1^2 m_2^2})$. Hence, if we face the unfortunate situation that both stops are light and $hGG$ deviations are absent due to a pernicious cancellation between stop loop contributions to the $hGG$ coupling, it may still be possible to indirectly constrain this scenario through di-Higgs production measurements at the LHC.

\subsection*{Additional Indirect Constraints}
\label{sec:CCB}
As mentioned in Sec.~\ref{sec:intro}, there are other indirect constraints on stops, and here we briefly comment on how those compare to the constraints and predictions considered here. The most relevant of these constraints comes from the observation that much of the parameter space considered has very large $A$-terms, and this can generate charge- or color-breaking vacua in the scalar potential that are deeper than the electroweak vacuum~\cite{Nilles:1982dy,AlvarezGaume:1983gj,Derendinger:1983bz,Claudson:1983et,Kounnas:1983td,Drees:1985ie,Gunion:1987qv,Komatsu:1988mt,Langacker:1994bc,Casas:1995pd,Casas:1996de}. One can approximate the maximum allowed $A$-term by considering a $D$-flat direction in field space where $\langle H_u \rangle = \langle \tilde{t}_L \rangle = \langle \tilde{t}_R \rangle$, and requiring that all minima in that direction have positive vacuum energy. This leads to the condition~\cite{Nilles:1982dy,AlvarezGaume:1983gj,Derendinger:1983bz,Claudson:1983et,Kounnas:1983td}
\begin{equation}
A_t^2 < 3 \left( m_{H_u}^2 + |\mu|^2 + m_{Q_3}^2+m_{U_3}^2 \right) \, .
\label{eq:vac-stab}
\end{equation}
In the decoupling limit, $m_{H_u}^2 + |\mu|^2 = - m_h^2/2$ where $m_h$ is the physical Higgs mass. We can take the small $\mu$ or large $\tan\beta$ limit which sets $A_t = X_t$. This allows us to plot the bound from \eref{vac-stab} in \Fig{fig:m1m2}, \Fig{fig:mA}, and \Fig{fig:mST}.

We stress that \eref{vac-stab} is a very crude approximation for the stability bound on the $A$-terms. In order to properly compute the bound, one must take into account loop contributions~\cite{Bordner:1995fh,Ferreira:2000hg}, tunneling effects~\cite{Riotto:1995am,Kusenko:1996jn,Kusenko:1996xt}, and properly account for cosmological history~\cite{Falk:1996zt}. There now exist sophisticated computer codes~\cite{Camargo-Molina:2013qva} which can compute bounds in various different supersymmetric models~\cite{Camargo-Molina:2013sta,Camargo-Molina:2014pwa,Chamoun:2014eda}. Other groups have recently considered the effect of the 125 GeV Higgs on these bounds~\cite{Chowdhury:2013dka,Blinov:2013fta,Bobrowski:2014dla,Chattopadhyay:2014gfa}. A full computation of the vacuum stability of our scenario is beyond the scope of this work, so we use \eref{vac-stab} to give a rough sense of where those bounds would lie.

Precision electroweak observables can also be used to constrain this scenario~\cite{Heinemeyer:2004gx}. One particularly important constraint comes from $\rho$-parameter which measures the splitting of electroweak multiplets. This constraint depends sensitively on the mass of the right handed sbottom as well as the mixing in the sbottom sector, and because of this additional model dependance we do not show the constraint on our figures. We find that generically the constraints from the $\rho$-parameter are weaker than the vacuum stability constraints.

\section{Conclusions and Outlook}
\label{sec:conclusion}

The observation and study of the di-Higgs channel is a primary objective of the LHC as well as future hadron colliders and it is a promising place to look for signatures of BSM physics. In this paper we have explored the impact of new colored states coupled to the Higgs particle on the production of Higgs boson pairs. Such states are well motivated by naturalness, with prime examples being top-partners. This class of non-resonant new physics can in principle lead to significant modifications to di-Higgs production. In most cases, however, the current experimental constraints on single Higgs production in the gluon fusion channel limit the extent to which the di-Higgs rate can deviate from the SM prediction. This can easily be seen in the case of heavy new colored states from an EFT analysis. The case of new light colored states requires a more detailed specification of the model and a full one loop calculation of the di-Higgs rate. We have performed such an analysis for the case of stops in supersymmetry, finding that that modifications are typically small, but that tuned regions with $\cO(1)$ enhancements to the cross sections exist. 

This result demonstrates that future di-Higgs measurements could be used to place indirect constraints on the presence of light stops if they have somehow otherwise evaded detection at the LHC. However, these modifications are likely to be modest given the present constraints on single Higgs production. Thus, if large modifications in the di-Higgs production rate were observed this work would suggest that they are more likely to come from resonant new physics, or modifications of the weak sector and/or Higgs self-coupling, rather than from non-resonant contributions from new colored fields coupled to the Higgs.

\acknowledgments
We would like to thank Zackaria Chacko, Matt Dolan, Ian Low, Andreas Papaefstathiou, Michael Spannowsky, and Michael Spira for helpful conversations. C.V. is supported by the National Science Foundation Grant No. PHY-1315155 and the Maryland Center for Fundamental Physics. B.B. and M.M are supported by CERN COFUND Fellowships.


\appendix

\section{100 TeV Projections}
\label{sec:100TeV}
We here provide estimates for the magnitude of corrections and expected sensitivity at a 100 TeV collider. The fractional cross section corrections are similar to the corrections at 14 TeV, however we have chosen to show more optimistic contours for the expected sensitivity due to increased overall cross sections and hence improved statistical uncertainties at 100 TeV, thus the plots look somewhat different.\footnote{Presumably systematic errors would also improve by the time of 100 TeV operation, particularly in theory uncertainties.}

\begin{figure}[t]
\centering
\includegraphics[height=2.6in]{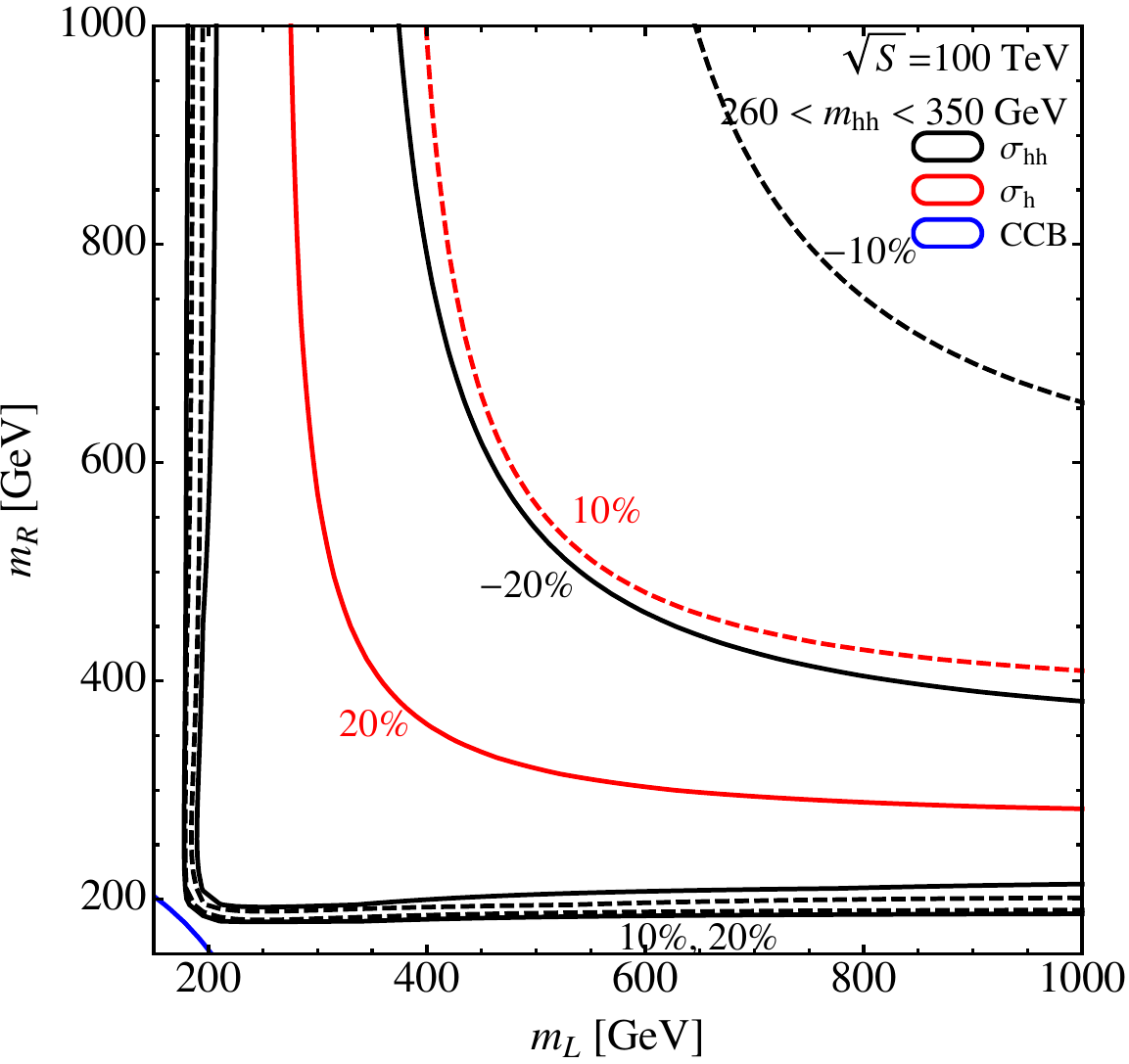}  \qquad \includegraphics[height=2.6in]{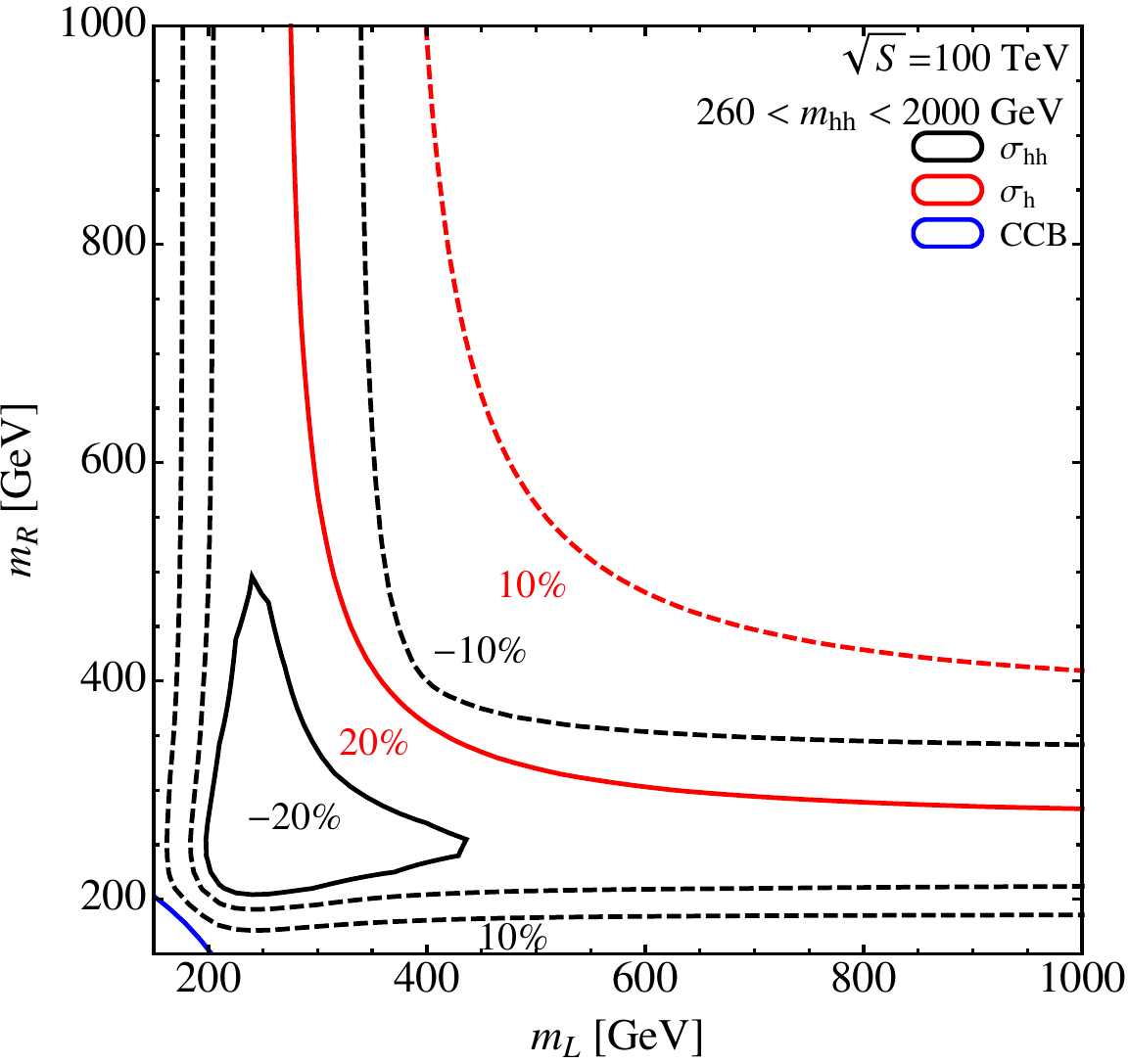}
\caption{Same as \Fig{fig:m1m2} but for $\sqrt{S} = 100$ TeV. }
\label{fig:m1m2100TeV}
\end{figure}

\begin{figure}[th!]
\centering
\includegraphics[height=2.6in]{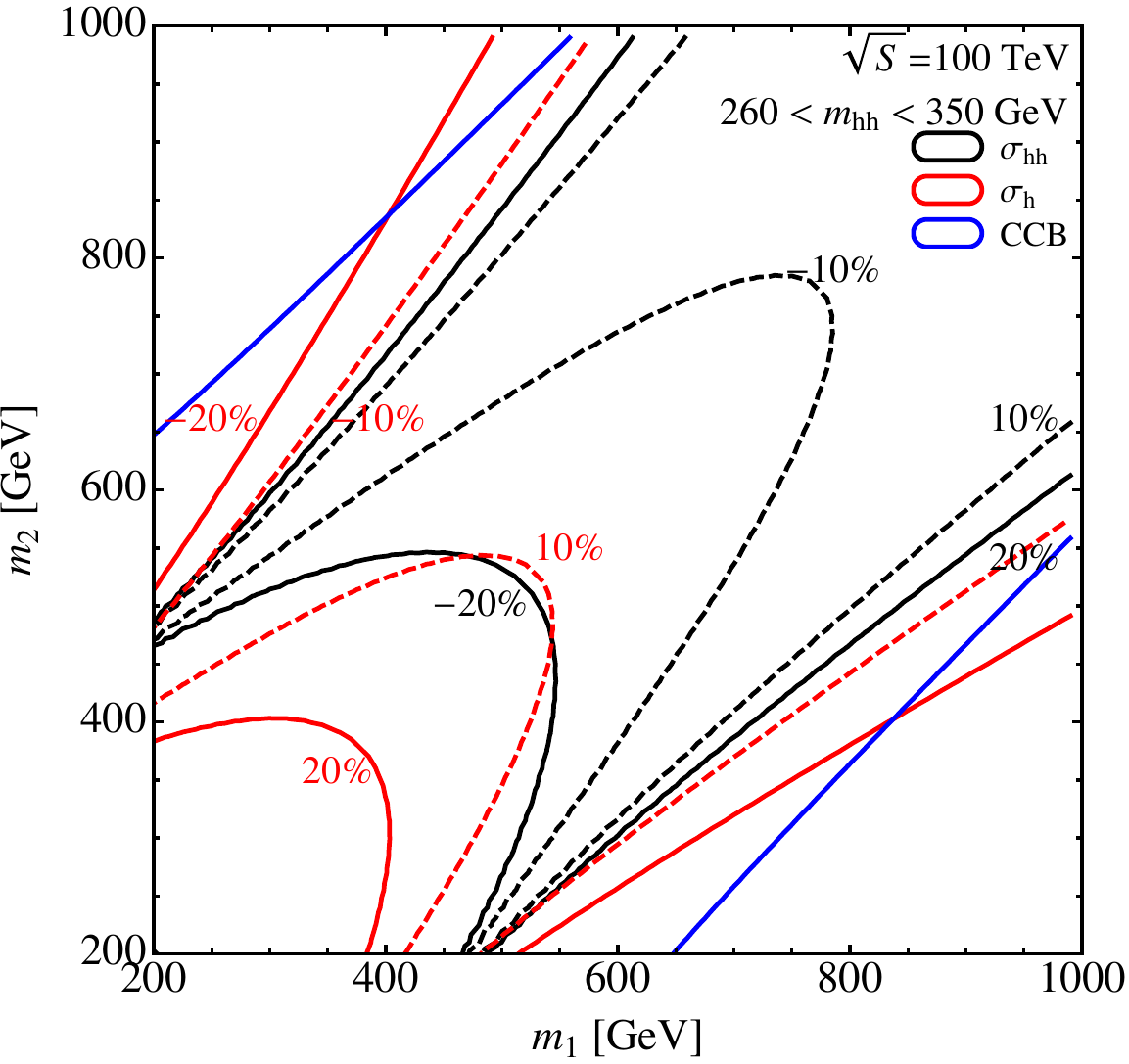} \qquad \includegraphics[height=2.6in]{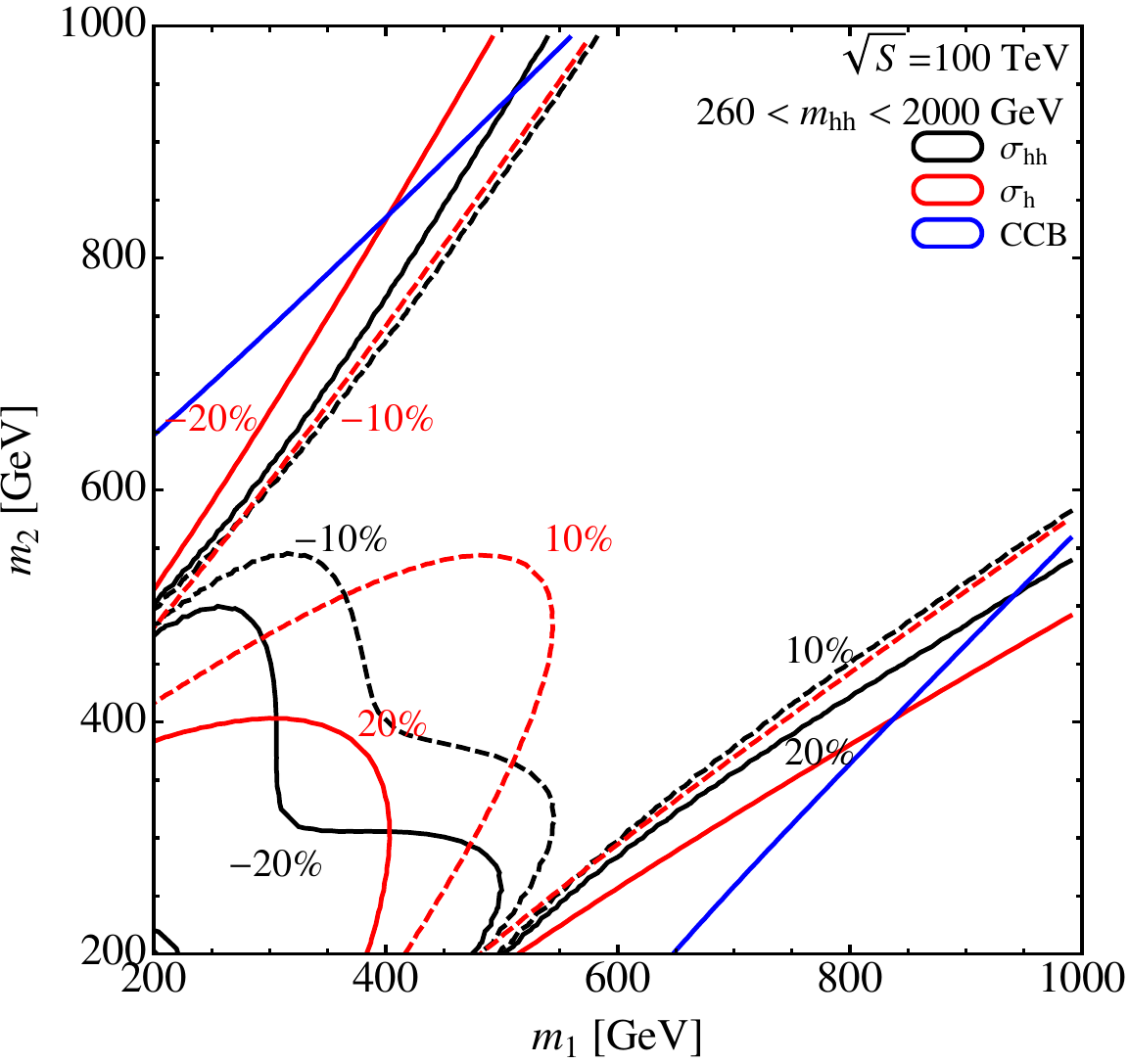}
\caption{Same as \Fig{fig:mA} but for $\sqrt{S} = 100$ TeV.}
\label{fig:mA100TeV}
\end{figure}

\begin{figure}[th!]
\centering
\includegraphics[height=2.6in]{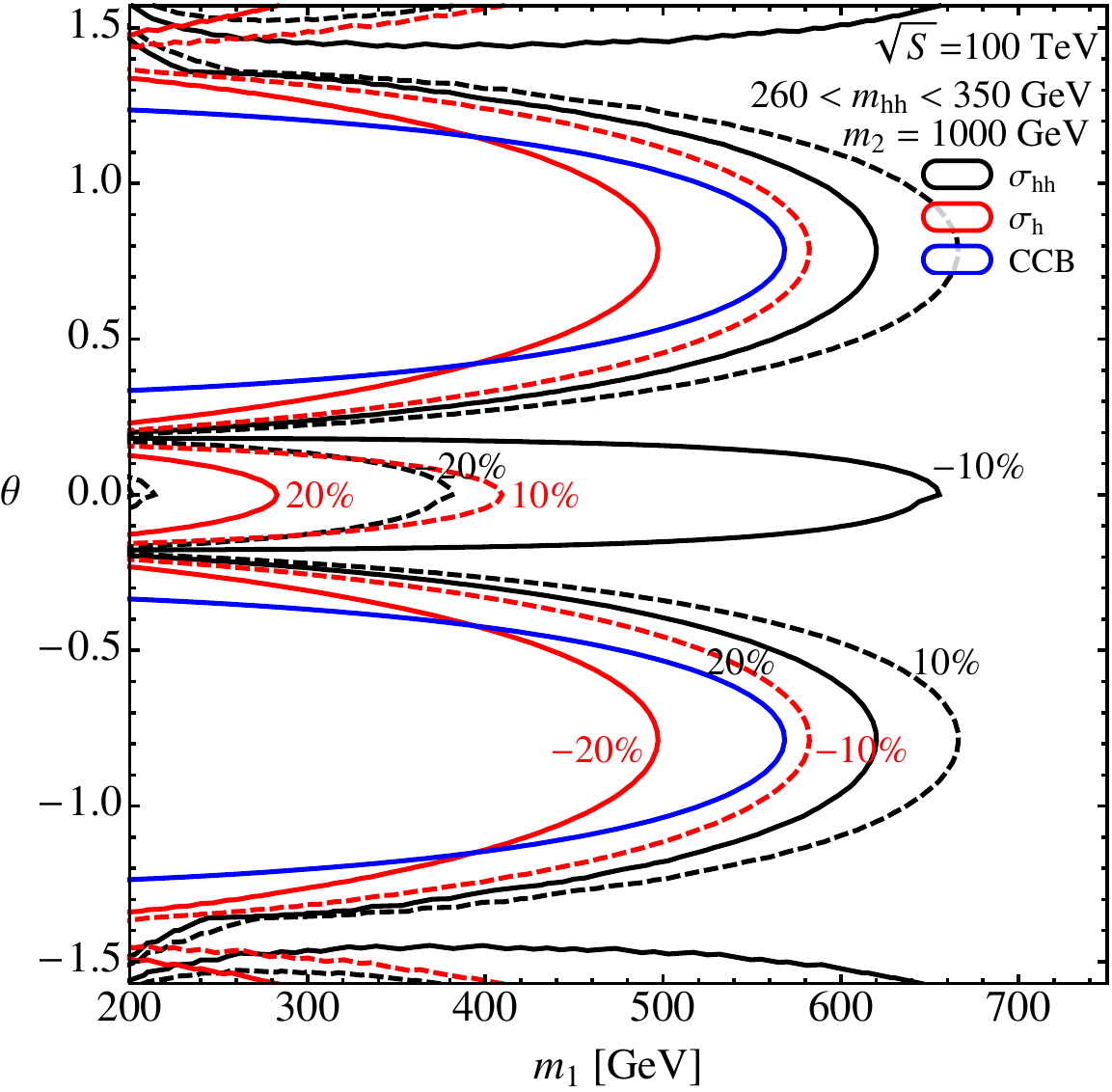} \qquad \includegraphics[height=2.6in]{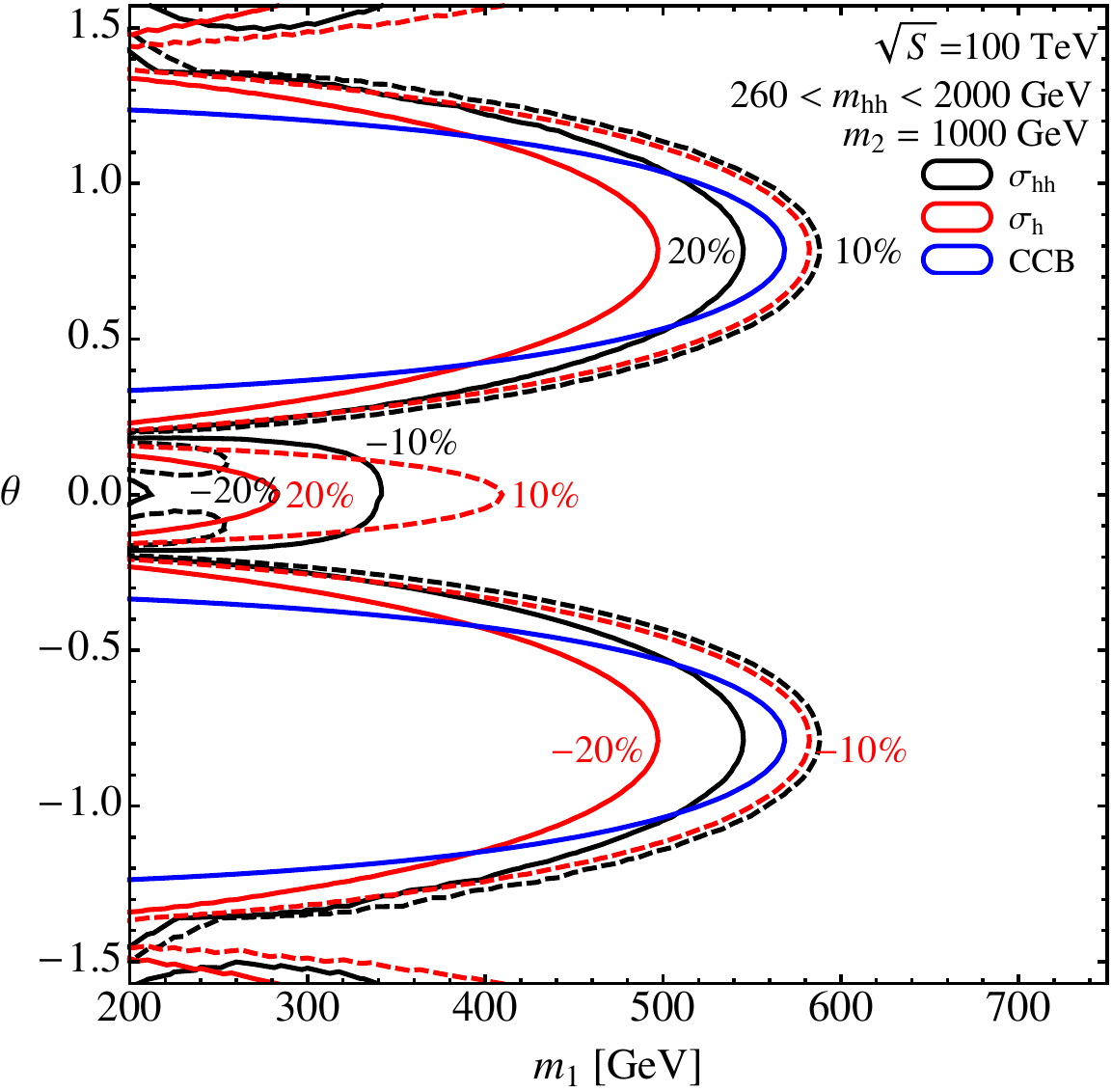}  \\ 
\includegraphics[height=2.6in]{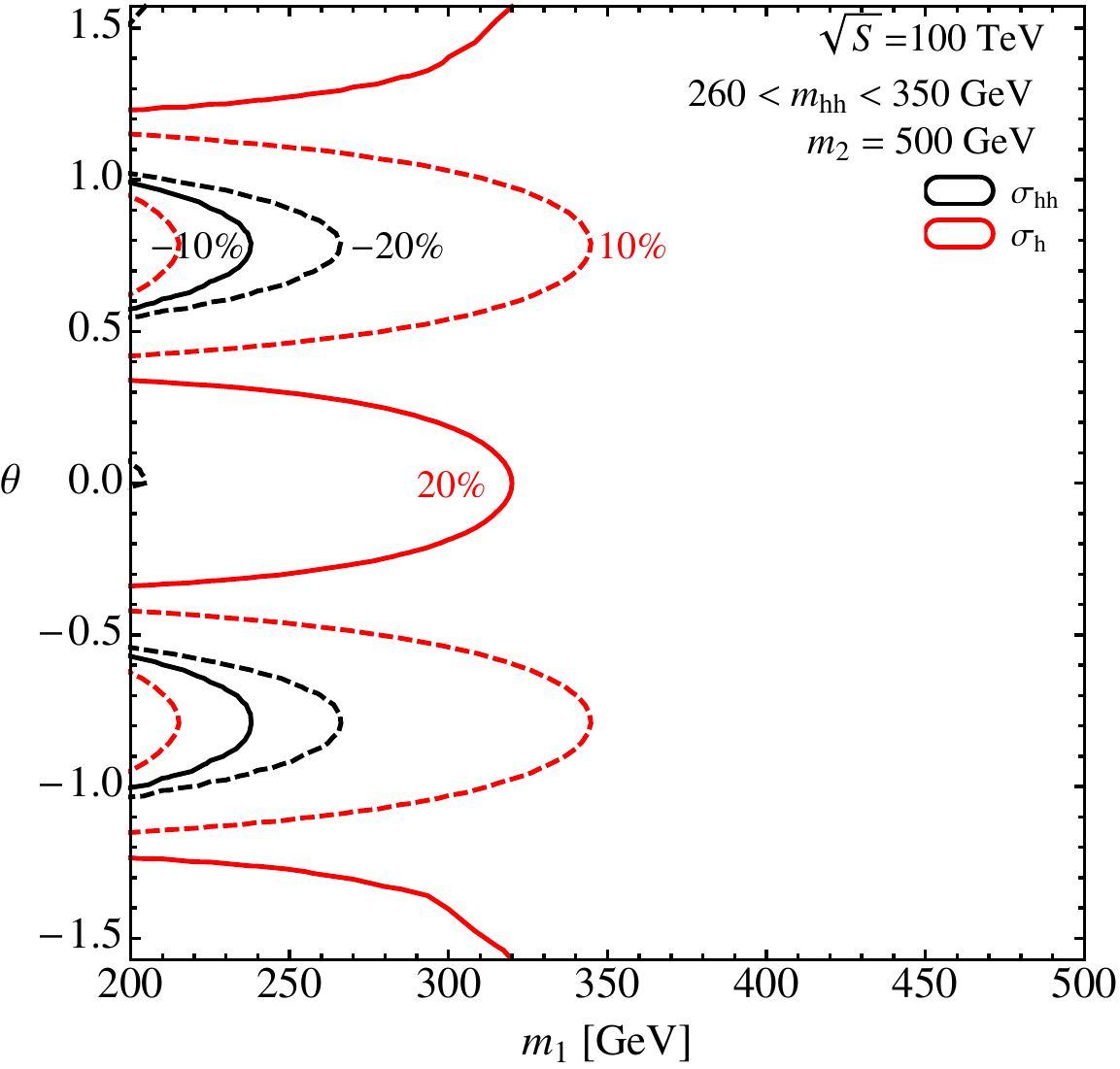} \qquad \includegraphics[height=2.6in]{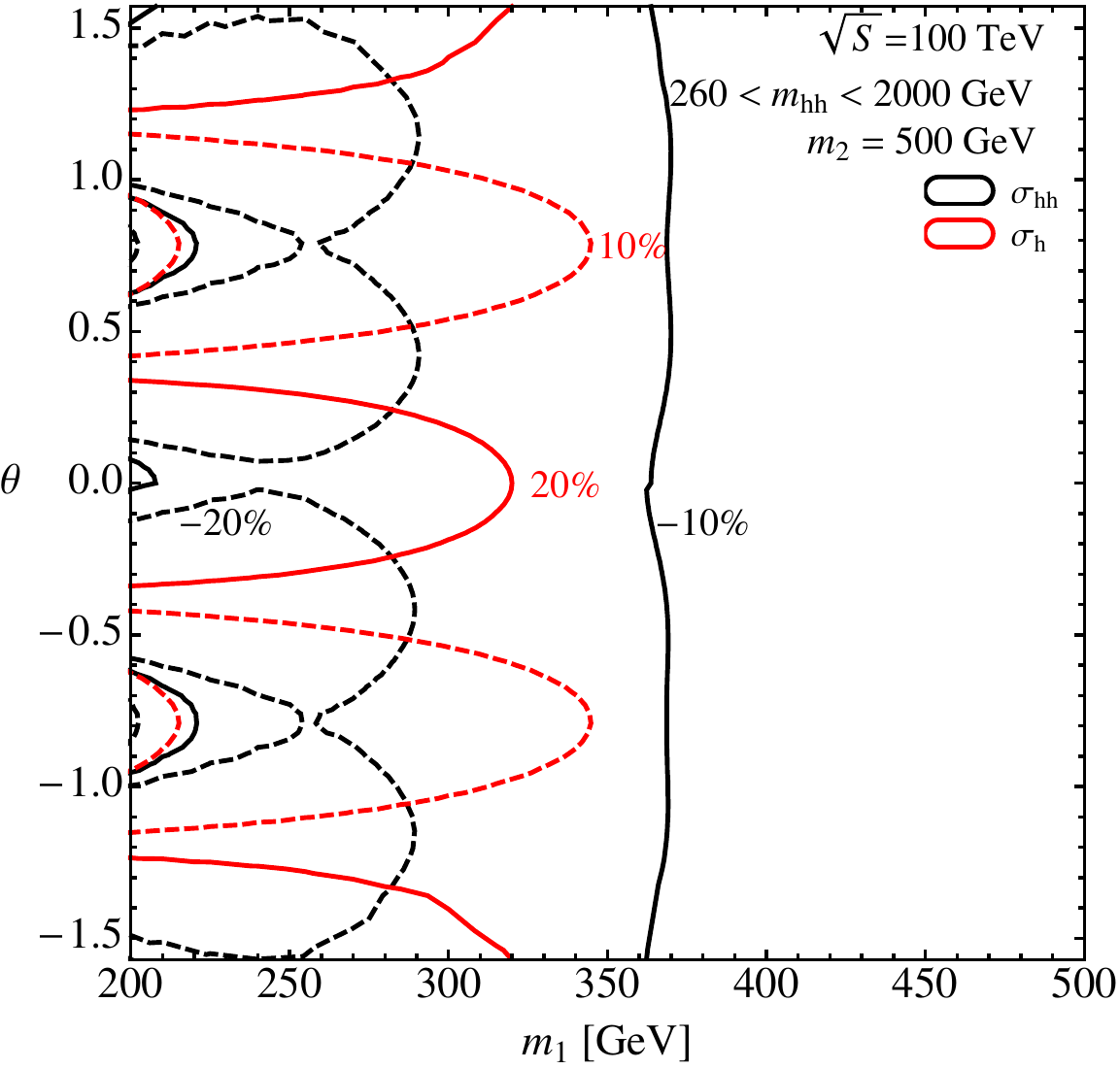} 
\caption{Same as \Fig{fig:mST} but for $\sqrt{S} = 100$ TeV.}
\label{fig:mST100TeV}
\end{figure}

\bibliographystyle{apsrev}
\bibliography{lit}

\end{document}